\date{\today}

\documentclass[12pt]{article}
\usepackage{amsmath}
\usepackage{graphicx}
\usepackage{lineno,hyperref}
\usepackage{multirow}
\usepackage[margin=1in]{geometry}
\usepackage[utf8]{inputenc}

\usepackage[ruled,vlined,linesnumbered]{algorithm2e}
\usepackage[style=authoryear, maxbibnames=99, mincitenames=1, maxcitenames=3, backref=true, hyperref=true, dashed=false, giveninits=true, backend=biber, bibencoding=utf8, uniquename=false, uniquelist=false]{biblatex}
\usepackage{float}
\usepackage{apatitlepage}
\usepackage{amsthm}
\usepackage[title,titletoc]{appendix}
\usepackage{bm}
\usepackage[T1]{fontenc}
\renewbibmacro{in:}{}
\newtheorem{assumption}{H}
\SetAlFnt{\small\sffamily}
\let\oldnl\nl% Store \nl in \oldnl
\newcommand{\nonl}{\renewcommand{\nl}{\let\nl\oldnl}}% Remove line number for one line

\newtheorem{theorem}{Theorem}
\threeauthors{Amalan Mahendran}{Helen Thompson}{James M. McGree}
\threeaffiliations{School of Mathematical Sciences\\ and Centre for Data Science, \\
Queensland University of Technology}{School of Mathematical Sciences\\ and Centre for Data Science, \\
Queensland University of Technology}{School of Mathematical Sciences\\ and Centre for Data Science, \\
Queensland University of Technology}
\APAtitle{A model robust sub-sampling approach for \\ Generalised Linear Models in Big data settings}
\APAshorttitle{ Model robust sub-sampling for Big data}
\APAauthornote{amalan.mahendran@qut.edu.au}
\APAkeywords{Large data set, model averaging, optimal sub-sampling}

\APAabstract{In today's modern era of Big data, computationally efficient and scalable methods are needed to support timely insights and informed decision making.
One such method is sub-sampling, where a subset of the Big data is analysed and used as the basis for inference rather than considering the whole data set.
A key question when applying sub-sampling approaches is how to select an informative subset based on the questions being asked of the data.
A recent approach for this has been proposed based on determining sub-sampling probabilities for each data point, but a limitation of this approach is that appropriate sub-sampling probabilities rely on an assumed model for the Big data.
In this article, to overcome this limitation, we propose a model robust approach where a set of models is considered, and the sub-sampling probabilities are evaluated based on the weighted average of probabilities that would be obtained if each model was considered singularly.
Theoretical support for such an approach is provided.
Our model robust sub-sampling approach is applied in a simulation study and in two real world applications where performance is compared to current sub-sampling practices.
The results show that our model robust approach outperforms alternative approaches.
}

\renewcommand{\APAtitleformat}{\LARGE\bfseries}

\DeclareMathOperator*{\argmaxA}{arg\,max}

\begin{document}
\APAmaketitle
\clearpage

\section{Introduction}\label{Sec:Introduction}

Big data analysis has gained interest in recent years through providing new insights and unlocking hidden knowledge in different fields of study \parencite{karmakar2018statistical} including medicine \parencite{rehman2021leveraging}, fraud detection \parencite{vaughan2020efficient}, the oil and gas industry \parencite{nguyen2020systematic} and astronomy \parencite{zhang2015astronomy}.
However, the analysis of Big data can be challenging for traditional statistical methods and standard computing environments \parencite{wang2016statistical}. 
\textcite{martinez2020modeling} discuss storage and modelling solutions when handling such a large amount of data. 
In general, modelling solutions can be grouped into three broad categories: 1) \emph{sub-sampling methods}, where the analysis is performed on an informative sub-sample obtained from the Big data \parencite{kleiner2014scalable,ma2015statistical,ma2015leveraging,drovandi2017principles,wang2018logistic,wang2019linear,yao2019softmax,ai2020quantile,cheng2020IBOSSlogistic,lee2021fast,ai2021optimal,yao2021review}; 2) \emph{divide and recombine methods}, where the Big data set is divided into smaller blocks, and then the intended statistical analysis is performed on each block and subsequently recombined for inference \parencite{lin2011aggregated,guha2012large,cleveland2014divide,chang2017divide,li2020sequential}; 3) \emph{online updating of streamed data}, where statistical inference is updated as new data arrive sequentially \parencite{schifano2016online,xue2020online}.   
In recent years, compared to divide and recombine methods, sub-sampling has been applied to a variety of regression problems, while online updating is typically only used for streaming data.
In addition, in cases where a large data set is not needed to answer a specific question with sufficient confidence, sub-sampling seems preferable as the analysis of the data can often be undertaken with standard methods.
Moreover, the computational efficiency of sub-sampling over analysing large data sets has been observed for parameter estimation in linear \parencite{wang2019linear} and logistic \parencite{wang2018logistic} regression models.
For these reasons, we focus on sub-sampling methods in this article.

The key challenge for sub-sampling methods is how to obtain an informative sub-sample that can be used to efficiently answer specific analysis questions and provide results that align with the analysis of the whole Big data set. 
Two approaches for this exist in the literature: 1) randomly sample from the Big data with sub-sampling probabilities that are found based on a specific statistical model and objective (e.g., prediction, parameter estimation) \parencite{wang2018logistic,yao2019softmax,wang2019linear,ai2020quantile,cheng2020IBOSSlogistic,ai2021optimal,lee2021fast,yao2021review}; 2) select sub-samples based on an experimental design \parencite{drovandi2017principles,Laura2022Optimal}. 
Randomly sampling with certain probabilities (based upon the definitions of $A$- or $L$- optimality criteria, see \textcite{atkinson2007optimum}) is the focus of this article, and has been applied for parameter estimation in a wide range of regression problems including softmax \parencite{yao2019softmax} and quantile regression \parencite{ai2020quantile}, and generalised linear models \parencite{wang2018logistic,wang2019linear,cheng2020IBOSSlogistic,ai2021optimal,yao2021review}.
In contrast, the approach based on an experimental design has only been applied for: 1) parameter estimation in logistic fixed and mixed effects regression models \parencite{drovandi2017principles}; 2) parameter estimation and prediction accuracy in linear and logistic regression models \parencite{Laura2022Optimal}.

A key feature of both of the current sub-sampling approaches is that they rely on a statistical model that is assumed to appropriately describe the Big data.
Given this is a potentially limiting assumption, \textcite{yu2022subdata} proposed to select the best candidate model from a pool of models based on the Bayesian Information Criterion (BIC) \parencite{schwarz1978estimating}.
This was applied to linear models, and resulted in sub-sampling probabilities that were more appropriate than those based on considering a single model.
Similarly, for linear regression, \textcite{shi2021model,meng2021lowcon} explored using space filling and orthogonal Latin hypercube experimental design techniques to allow for potential model misspecification. 
In this paper, we propose that, instead of selecting a single best candidate model for the Big data, a set of models is considered and a model averaging approach is used for determining the sub-sampling probabilities.
Through adopting such an approach, it is thought that the analysis goal (e.g., efficient parameter estimation) should be achieved regardless of the preferred model for the data. 
To implement this model robust approach, we consider sub-sampling based on $A$- and $L$- optimality criteria within the Generalised Linear Modelling framework, and provide theoretical support for using a model averaged approach based on each of these criteria.
Given we consider Generalised Linear Models (GLMs), our approach should be generally applicable across many areas of science and technology where a variety of data types are observed.
This is demonstrated through applying our proposed methods within a simulation study, and for the analysis of two real-world Big data problems.

The remainder of the article is structured as follows. 
Section~\ref{Sec:Background} introduces GLMs and the existing probability-based sub-sampling approach of \textcite{ai2021optimal} and \textcite{yao2021review}.
Our proposed model robust sub-sampling approach is introduced in Section~\ref{Sec:ModRobOptSubMethod}, which is embedded with a GLM framework.
A simulation study is then used to assess the performance of our model robust approach in Section~\ref{Sec:SimulationAndRealWorldSetup}, and two real world applications are presented. 
Section~\ref{Sec:Discussion} concludes the article with a discussion of the results and some suggestions for future research. 

\section{Background}\label{Sec:Background}

There are variety of ways Big data can be sub-sampled.  
In this section, we focus on the approach where sub-sampling probabilities are determined for each data point, and the Big data is sub-sampled (at random) based on these probabilities. 
Such an approach was first proposed by \textcite{wang2018logistic} for logistic regression problems in Big data settings, and has been extended to a wide range of regression problems (e.g., \textcite{yao2019softmax,wang2019linear,ai2020quantile,cheng2020IBOSSlogistic,ai2021optimal,yao2021review}). 
In this section, we describe such a sub-sampling approach as applied to GLMs based on the work of \textcite{ai2021optimal}.

\subsection{Generalised Linear Models}\label{Sec:GLMs}

Let a Big data set be denoted as $F_N=(\bm{X}_0,\bm{y})$, where $\bm{X}_0=({\bm{x}_0}_1,\ldots,{\bm{x}_0}_N)^T \in R^{N \times p}$ represents a data matrix based on the Big data set with $p$ covariates, $\bm{y}=(y_1,\ldots,y_N)^T$ represents the response vector and $N$ is the total number of data points.
To fit a GLM, consider the model matrix $\bm{X}=h(\bm{X}_0) \in R^{N \times (p+q)}$ where $h(.)$ is some function of $\bm{X}_0$ which creates an additional $q$ columns representing, for example, an intercept and/or higher-order terms.
A GLM can then be defined via three components: 1) distribution of response $\bm{y}$, which is from the exponential family (e.g., Normal, Binomial or Poisson); 2) linear predictor $\bm{\eta}=\bm{X}\bm{\theta}$, where $\bm{\theta}=(\theta_1,\ldots,\theta_{p+q})^T$ is the parameter vector; and 3) link function $g(.)$, which links the mean of the response to the linear predictor \parencite{nelder1972generalized}.
Throughout this article, the inverse link function $g^{-1}(.)$ is denoted by $u(.)$.

A common exponential form for the probability density or mass function of $y$ can be written as:
\begin{equation}\label{Eq:ypdf}
    f(y;\omega,\gamma)=\exp{\Big(\frac{y\omega - \psi(\omega)}{a(\gamma)} + b(y,\gamma) \Big)},
\end{equation}
where $\psi(.), a(.)$ and $b(.)$ are some functions, $\omega$ is known as the natural parameter and $\gamma$ the dispersion parameter. 
Based on Equation \eqref{Eq:ypdf} the link function $g(.)$ can then be defined as $g(\bm{\mu}) = \bm{\eta}$, where $\mu = E[y|\bm{X},\bm{\theta}] = \mbox{d} \psi(\omega)/\mbox{d}\omega$.
A general linear model, or linear regression model, is a special case of a GLM where $\bm{y} \sim N(\bm{\mu},\bm{\Sigma})$ and $g(.)$ is the identity link function $g(\bm{\mu})=\bm{\mu}$, such that $\bm{\mu}=\bm{X}\bm{\theta}$. 
For logistic regression, $\bm{y} \sim \mbox{Bin}(n,\bm{\pi})$ and $g(.)$ is the logit link function $g(\bm{\pi})=\log(\bm{\pi}/(1-\bm{\pi}))$ such that $g(\bm{\pi})=\bm{X}\bm{\theta}$. Similarly, for Poisson regression, $\bm{y} \sim \mbox{Poisson}(\bm{\lambda})$ and $g(.)$ is the log link function $g(\bm{\lambda})=\log(\bm{\lambda})$ such that $g(\bm{\lambda})=\bm{X}\bm{\theta}$. 
Note that the dispersion parameter $\gamma=1$ for the logistic and Poisson regression models.

\subsection{A general sub-sampling algorithm for GLMs}\label{Sec:GenSamAlgoGLMs}

As described by \textcite{ai2021optimal}, consider a general sub-sampling approach to estimate parameters $\bm{\theta}$ through a weighted log-likelihood function for GLMs (weights are the inverse of the sub-sampling probabilities). 
A weighted likelihood function is considered, since an unweighted likelihood leads to biased estimates of model parameters for logistic regression, see \textcite{wang2019more}.
Define $\phi_i$ as the probability that row $i$ of $F_N$ is randomly selected, for $i=1,\ldots,N$, where $\sum_{i=1}^{N} \phi_i=1$ and $\phi_i \in (0,1)$.
A sub-sample $S$ of size $r$ is then drawn with replacement from $F_N$ based on $\bm{\phi}=(\phi_1,\ldots,\phi_N)$.
The selected responses, covariates and sub-sampling probabilities are then used to estimate the model parameters.   
Pseudo-code for this general sub-sampling approach is provided in Algorithm~\ref{Algo:GenSam}.
\begin{algorithm}[htbp!]
\SetAlgoLined
    \textbf{Sampling:} Assign $\phi_i,i=1,...,N,$ for $F_N$. \\ \nonl 
    Based on $\bm{\phi}$, draw an $r$ size sub-sample with replacement from $F_N$ to yield $S = \{\bm{x}_l^*,y_l^*,\phi_l^*\}_{l=1}^r = (\bm{X}^*,\bm{y}^*,\bm{\phi}^*)$. \\
    \textbf{Estimation:} Based on $S$, find:
    \begin{equation*}
    \begin{aligned}
        \tilde{\bm{\theta}} = \argmaxA_{\bm{\theta}} ~ \log{L(\bm{\theta}|\bm{X}^*,\bm{y}^*,\bm{\phi}^*)}  \equiv \argmaxA_{\bm{\theta}}~ \frac{1}{r}\sum_{l=1}^{r} \frac{y^*_l u(\bm{\theta}^T\bm{x}^*_l) - \psi(u(\bm{\theta}^T\bm{x}^*_l))}{\phi^*_l}.
    \end{aligned}
    \end{equation*} \\
    \textbf{Output:} $\tilde{\bm{\theta}}$ and $S$.
\caption{General sub-sampling algorithm \parencite{ai2021optimal}} \label{Algo:GenSam}
\end{algorithm}

From Algorithm~\ref{Algo:GenSam}, the first step is to assign sub-sampling probabilities $\phi_i$ to the rows of $F_N$.  
The simplest approach is to assign each data point an equal probability of being selected. 
These probabilities could also depend on the composition of $\bm{y}$, e.g., for binary data, one could sample proportional to the inverse of the number of successes and failures.  
Based on these probabilities, a sub-sample $S$ of size $r$ is then drawn completely at random (with replacement) from $F_N$ to yield $\{\bm{x}^*_l,y^*_l,\phi^*_l\}_{l=1}^r$.
Based on this sub-sample, model parameters $\tilde{\bm{\theta}}$ are estimated via maximising a weighted log-likelihood function.
The estimates $\tilde{\bm{\theta}}$ can then be considered as estimates of what would be obtained if the whole Big data set were analysed.

The asymptotic properties of $\tilde{\bm{\theta}}$ based on the general sub-sampling approach given in Algorithm~\ref{Algo:GenSam} were derived by \textcite{ai2021optimal}.  
These properties are outlined below as they form the basis for our extensions to model robust sub-sampling detailed in Section~\ref{Sec:ModRobOptSubMethod}.

\subsection[Asymptotic properties of \texorpdfstring{$\tilde{\bm{\theta}}$}  from the general sub-sampling algorithm]{Asymptotic properties of $\tilde{\bm{\theta}}$  from the general sub-sampling algorithm} \label{Sec:Asymptotic}

To explore the asymptotic properties of the estimates obtained from the general sub-sampling algorithm, \textcite{ai2021optimal} made a number of assumptions which are outlined below.  
To follow these, note that $\dot{\psi}(u(\bm{\theta}^T\bm{x}_i))$ and $\dot{u}(\bm{\theta}^T\bm{x}_i)$ denote the first-order derivatives of $\psi(u(\bm{\theta}^T\bm{x}_i))$ and $u(\bm{\theta}^T\bm{x}_i)$ (with respect to $\bm{\theta}$), respectively, with two dots similarly denoting the second-order derivatives.
In addition, the Euclidean norm of a vector $\bm{a}$ will be denoted as $||\bm{a}||=(\bm{a}^T\bm{a})^{1/2}$.
\begin{assumption}\label{Ass:1}
    Assume that $\bm{X}\bm{\theta}$ lies in the interior of a compact set $K \in \Omega$ almost surely.
\end{assumption}

\begin{assumption}\label{Ass:2}
    The regression coefficient $\bm{\theta}$ is an inner point of the compact domain $\Lambda_B=\{\bm{\theta} \in R^p : ||\bm{\theta}|| \leq B \}$ for some constant $B$.
\end{assumption}

\begin{assumption}\label{Ass:3}
    Central moments condition: $N^{-1}\sum_{i=1}^{N}|y_i-\dot{\psi}(u(\bm{\theta}^T\bm{x}_i))|^4 = O_P(1)$ for all $\bm{\theta} \in \Lambda_B$.
\end{assumption}

\begin{assumption}\label{Ass:4}
    As $N \rightarrow \infty$, the observed information matrix 
    $$ \bm{J_X} := \frac{1}{N} \sum_{i=1}^{N} \{\ddot{u}(\hat{\bm{\theta}}^T_{MLE}\bm{x}_i) \bm{x}_i \bm{x}^T_i[\dot{\psi}(u(\hat{\bm{\theta}}^T_{MLE}\bm{x}_i)) - y_i] + \ddot{\psi}(u(\hat{\bm{\theta}}^T_{MLE}\bm{x}_i))\dot{u}^2(\hat{\bm{\theta}}^T_{MLE}\bm{x}_i)\bm{x}_i \bm{x}^T_i\}$$ 
    goes to a positive definite matrix in probability.
\end{assumption}

\begin{assumption}\label{Ass:5}
    Require that the full sample covariates have finite $6$-th order moments, \,i.e., $E||\bm{x}_1||^6 \leq \infty$.
\end{assumption}

\begin{assumption}\label{Ass:6}
    Assume $N^{-2}\sum_{i=1}^{N} ||\bm{x}_i||^s/ \phi_i = O_P(1)$ for $s=2,4$.
\end{assumption}

\begin{assumption}\label{Ass:7}
    For $\delta=0$ and some $\delta > 0$, assume 
    $$\frac{1}{N^{2+\delta}} \sum_{i=1}^{N} \frac{|y_i - \dot{\psi}_i(u(\hat{\bm{\theta}}^T_{MLE}\bm{x}_i))|^{2+\delta} ||\dot{u}(\hat{\bm{\theta}}^T_{MLE}\bm{x}_i)\bm{x}_i||^{2+\delta}}{\phi^{1+\delta}_i} = O_P(1).$$
\end{assumption}
Assumptions H\ref{Ass:1} and H\ref{Ass:2} ensure that $\mbox{E}[y_i|\bm{x}_i] < \infty$ for all $i$, and was used by \textcite{clemenccon2014scaling} when investigating the impact of survey sampling with unequal inclusion probabilities on (stochastic) gradient descent-based-estimation methods in Big data problems. 
H\ref{Ass:2} defines an admissible set which is required for ensuring consistency of estimates for model parameters for GLMs, see \textcite{fahrmeir1985consistency}.
Assumption H\ref{Ass:4} ensures (asymptotically) that $\bm{J_X}$, the observed Fisher information, is defined for the given model, and that it is positive definite and therefore non-singular.
To obtain a Bahadur representation, that is often useful in determining the asymptotic properties of statistical estimators of the sub-sampled estimator, Assumptions H\ref{Ass:3} and H\ref{Ass:5} are needed.
H\ref{Ass:6} and H\ref{Ass:7} are moment conditions on covariates and sub-sampling probabilities. 
Assumption H\ref{Ass:7} is required by the Lindeberg-Feller central limit theorem, see \textcite{van2000asymptotic}.

Based on the above assumptions, the following theorems were proved by \textcite{ai2021optimal}. 
The first theorem proves that the estimator from the sub-sampling algorithm is consistent where $||\tilde{\bm{\theta}} - \hat{\bm{\theta}}_{MLE}||$ can be made small with a large sub-sample size $r$, and $\hat{\bm{\theta}}_{MLE}$ is the maximum likelihood estimator of $\bm{\theta}$ based on $F_N$.
The second theorem proves that the approximation error, $\tilde{\bm{\theta}} - \hat{\bm{\theta}}_{MLE}$, given $F_N$ is approximately asymptotically Normally distributed with mean zero and variance $\bm{V}$.
\begin{theorem}\label{The:1}
    If H\ref{Ass:1} to H\ref{Ass:7} hold then, as $N \rightarrow \infty$ and $r \rightarrow \infty$, $\tilde{\bm{\theta}}$ is consistent to $\hat{\bm{\theta}}_{MLE}$ in conditional probability given $F_N$. Moreover, the rate of convergence is $r^{-\frac{1}{2}}$. That is, with probability approaching one, for any $\epsilon > 0$, there exist finite $\Delta_{\epsilon}$ and $r_{\epsilon}$ such that 
    \begin{equation*}
        P(||\tilde{\bm{\theta}} - \hat{\bm{\theta}}_{MLE}|| \ge r^{-\frac{1}{2}} \Delta_{\epsilon} | F_N ) < \epsilon, \forall r > r_{\epsilon}.
    \end{equation*}
\end{theorem}
\begin{theorem}\label{The:2}
    If H\ref{Ass:1} to H\ref{Ass:7} hold, then as $N \rightarrow \infty$ and $r \rightarrow \infty$, conditional on $F_N$ in probability, 
    \begin{equation*}
        \bm{V}^{-1/2}(\tilde{\bm{\theta}}-\hat{\bm{\theta}}_{MLE}) \rightarrow N(\bm{0},\bm{I}),
    \end{equation*}
    in distribution, where $\bm{V}= \bm{J}^{-1}_{\bm{X}} \bm{V}_c \bm{J}^{-1}_{\bm{X}}= O_P(r^{-1})$ and 
    \begin{equation*}
        \bm{V}_c = \frac{1}{rN^2} \sum_{i=1}^{N} \frac{\{y_i - \dot{\psi}(u(\hat{\bm{\theta}}^T_{MLE}\bm{x}_i))\}^2 \dot{u}^2(\hat{\bm{\theta}}^T_{MLE}\bm{x}_i)\bm{x}_i \bm{x}^T_i}{\phi_i}.
    \end{equation*}
\end{theorem}
When applying the general sub-sampling algorithm, it may not be clear how to appropriately choose $\bm{\phi}$ depending upon the goal of the analysis (e.g., parameter estimation, response prediction, etc).
To address this, \textcite{ai2021optimal} proposed determining $\bm{\phi}$ based on an optimality criterion from experimental design (specifically $A$-optimality and $L$-optimality), and this led to the proposal of Theorems~\ref{The:3} and \ref{The:4} (below).
Such theorems provide the optimal choice for the sub-sampling probabilities to minimise the asymptotic mean squared error of $\tilde{\bm{\theta}}$ (or $\mbox{tr}(\bm{V})$) and $\bm{J_X} \tilde{\bm{\theta}}$ (or $\mbox{tr}(\bm{V}_c)$), denoted as $\bm{\phi}^{mMSE}$ and $\bm{\phi}^{mV_c}$, respectively. 
We note that such sub-sampling probabilities are conditional on the assumption of a model being appropriate to describe the Big data.
\begin{theorem}\label{The:3}
    The sub-sampling strategy is $A$-optimal if the sub-sampling probability is chosen such that 
    \begin{equation}
        \phi^{mMSE}_i = \frac{|y_i - \dot{\psi}(u(\hat{\bm{\theta}}^T_{MLE}\bm{x}_i))|\, ||\bm{J}^{-1}_{\bm{X}} \dot{u}(\hat{\bm{\theta}}^T_{MLE}\bm{x}_i)\bm{x}_i ||}{\sum_{j=1}^{N} |y_j - \dot{\psi}(u(\hat{\bm{\theta}}^T_{MLE}\bm{x}_j))|\, ||\bm{J}^{-1}_{\bm{X}} \dot{u}(\hat{\bm{\theta}}^T_{MLE}\bm{x}_j)\bm{x}_j ||}, i=1,\ldots,N.
    \end{equation}
\end{theorem}
Optimal sub-sampling probabilities from Theorem~\ref{The:3} can be computationally expensive as they require the calculation of $\bm{J}^{-1}_{\bm{X}}$.
Hence, the linear- or $L$- optimality criterion was proposed by \textcite{ai2021optimal}, which minimises the asymptotic mean squared error of $\bm{J_X}\tilde{\bm{\theta}}$ or $\mbox{tr}(\bm{J_X} \bm{V} \bm{J_X})$ which corresponds to minimising the variance of a linear combination of the parameters  \parencite{atkinson2007optimum}.  
This led to the following theorem.
\begin{theorem}\label{The:4}
    The sub-sampling strategy is $L$-optimal if the sub-sampling probability is chosen such that 
    \begin{equation}
        \phi^{mV_c}_i = \frac{|y_i - \dot{\psi}(u(\hat{\bm{\theta}}^T_{MLE}\bm{x}_i))|\, || \dot{u}(\hat{\bm{\theta}}^T_{MLE}\bm{x}_i)\bm{x}_i||}{
        \sum_{j=1}^{N} |y_j - \dot{\psi}(u(\hat{\bm{\theta}}^T_{MLE}\bm{x}_j))|\, ||\dot{u}(\hat{\bm{\theta}}^T_{MLE}\bm{x}_j)\bm{x}_j ||}, i=1,\ldots,N.
    \end{equation}
\end{theorem}

\textcite{wang2018logistic} applied Theorems~\ref{The:3} and \ref{The:4} to obtain the following optimal sub-sampling probabilities for logistic regression:
\begin{align}
        \phi^{mMSE}_i = \frac{|y_i - \pi_i|\,|| \bm{J}^{-1}_{\bm{X}} \bm{x}_i ||}{\sum_{j=1}^{N} |y_j - \pi_j|\,|| \bm{J}^{-1}_{\bm{X}} \bm{x}_j||}, & \quad \phi^{mV_c}_i = \frac{|y_i - \pi_i|\,||\bm{x}_i||}{\sum_{j=1}^{N} |y_j - \pi_j|\,||\bm{x}_j||}
\end{align}
with $y_i \in \{0, 1\}$, $\pi_i = \exp{(\hat{\bm{\theta}}^T_{MLE}\bm{x}_i)}/(1+\exp{(\hat{\bm{\theta}}^T_{MLE}\bm{x}_i)})$, $\bm{J_X}=N^{-1}\sum_{h=1}^{N} \pi_h(1-\pi_h)\bm{x}_h {\bm{x}_h}^T$ and $i=1,\ldots,N$.

\textcite{ai2021optimal} applied Theorems~\ref{The:3} and \ref{The:4} to obtain the following optimal sub-sampling probabilities for Poisson regression:
\begin{align}
    \phi^{mMSE}_i = \frac{|y_i - \lambda_i|\,|| \bm{J}^{-1}_{\bm{X}} \bm{x}_i ||}{\sum_{j=1}^{N} |y_j - \lambda_j|\,|| \bm{J}^{-1}_{\bm{X}} \bm{x}_j||}, & \quad \phi^{mV_c}_i = \frac{|y_i - \lambda_i|\,||\bm{x}_i||}{\sum_{j=1}^{N} |y_j - \lambda_j|\,||\bm{x}_j||} 
\end{align}
with $y_i \in N_0$ or non-negative integers, $\lambda_i = \exp{(\hat{\bm{\theta}}^T_{MLE}\bm{x}_i)}$, $\bm{J_X}=N^{-1}\sum_{h=1}^{N} \exp{(\hat{\bm{\theta}}^T_{MLE}\bm{x}_h)}\bm{x}_h {\bm{x}_h}^T$  and $i=1,\ldots,N$.

Unfortunately, in practice, the optimal sub-sampling probabilities $\bm{\phi}^{mMSE}$ and $\bm{\phi}^{mV_c}$ cannot be determined as they depend on $\hat{\bm{\theta}}_{MLE}$.
To address this, based on Theorems~\ref{The:1} and \ref{The:2}, \textcite{ai2021optimal} proposed a two stage sub-sampling strategy where an initial random sample of the Big data is used to estimate $\hat{\bm{\theta}}_{MLE}$; an estimate which is then used with the results from Theorems~\ref{The:3} and \ref{The:4} to provide estimates of optimal sub-sampling probabilities.  
Such an approach is thus termed a two stage sub-sampling approach, and is outlined in the next section.

\subsection{Optimal sub-sampling algorithm for GLMs}\label{Sec:OptSubAlgGLMs}

Theorems~\ref{The:1} to \ref{The:4} provide a theoretical basis for the optimal sub-sampling algorithm of \textcite{ai2021optimal}.
In this section, we describe this approach for GLMs which is outlined in Algorithm~\ref{Algo:OSGLMAC}.
This is a two stage algorithm where the general sub-sampling algorithm is initially applied.
For this, the sub-sampling probabilities could be $1/N$ for all observations or may be based on a stratified sampling technique.  
Based on this initial sub-sample, estimates $\tilde{\bm{\theta}}$ are obtained.
Then, optimal sub-sampling probabilities are estimated using results from Theorems~\ref{The:3} and \ref{The:4}, where $\hat{\bm{\theta}}_{MLE}$ is approximated by $\tilde{\bm{\theta}}$.

\begin{algorithm}[H]
\SetAlgoLined

    \nonl \textbf{Stage 1} \\ \vspace{0.1cm}
    \textbf{Random Sampling :} Assign $\bm{\phi} = (\phi_1,\ldots,\phi_N)$ for $F_N$. For example, in logistic regression $\phi_i=\phi^{prop}$ represents proportional sub-sampling probabilities that are based on the response composition, and $\phi_i=N^{-1}$ which is uniform sub-sampling probabilities for Poisson regression.\\ \nonl
    According to $\bm{\phi}$ draw a random sub-sample of size $r_0$, $S_{r_0}=\{\bm{x}^{r_0}_l,y^{r_0}_l,\phi^{r_0}_l\}_{l=1}^{r_0} = (\bm{X}^{r_0},\bm{y}^{r_0},\bm{\phi}^{r_0})$. \\
    \textbf{Estimation:} Based on $S_{r_0}$, find:
    \begin{equation*}
    \begin{aligned}
        \tilde{\bm{\theta}}^{r_0} = \argmaxA_{\bm{\theta}} ~ \log{L(\bm{\theta}|\bm{X}^{r_0},\bm{y}^{r_0},\bm{\phi}^{r_0})}  \equiv \argmaxA_{\bm{\theta}} ~ \frac{1}{r_0}\sum_{l=1}^{r_0} \Big[\frac{y^{r_0}_l u(\bm{\theta}^T\bm{x}^{r_0}_l) - \psi(u(\bm{\theta}^T\bm{x}^{r_0}_l))}{\phi^{r_0}_l} \Big].
    \end{aligned}
    \end{equation*} \nonl \\  
    \vspace{0.1cm} 
    \textbf{Stage 2} \\
    \vspace{0.1cm}
    \textbf{Optimal sub-sampling probability :} Estimate $\bm{\phi}^{mMSE}$ or $\bm{\phi}^{mV_c}$ from Theorems~\ref{The:3} and \ref{The:4}  using $\tilde{\bm{\theta}}^{r_0}$ in place of $\hat{\bm{\theta}}_{MLE}$. \\
    \textbf{Optimal sub-sampling and estimation:} Based on the estimated sub-sampling probabilities, draw a sub-sample $S_r$ completely at random (with replacement) of size $r$ from $F_N$, such that $S_{r}=\{\bm{x}^r_l,y^r_l,\phi^r_l\}_{l=1}^{r} = (\bm{X}^{r},\bm{y}^{r},\bm{\phi}^{r})$. \\ \nonl
    Form $S_{r_0+r}$ by combining $S_{r_0}$ and $S_{r}$, and obtain:
    \begin{align*}
        \tilde{\bm{\theta}} = & \argmaxA_{\bm{\theta}} ~ \log{L(\bm{\theta}|\bm{X}^{r_0},\bm{y}^{r_0},\bm{\phi}^{r_0},\bm{X}^{r},\bm{y}^{r},\bm{\phi}^{r})} \notag \\
        \equiv & \argmaxA_{\bm{\theta}} ~ \frac{1}{r_0+r} \Bigg[ \sum_{k \in \{r_0,r\}}  \sum_{l=1}^{k} \frac{y^{k}_l u(\bm{\theta}^T\bm{x}^{k}_l) - \psi(u(\bm{\theta}^T \bm{x}^{k}_l))}{\phi^{k}_l} \Bigg] .
    \end{align*} \\
    \textbf{Output :} $\tilde{\bm{\theta}}$ and $S_{r_0+r}$.
\caption{Two stage optimal sub-sampling algorithm for GLMs.}\label{Algo:OSGLMAC}
\end{algorithm}

The first stage of the algorithm is as shown in Algorithm~\ref{Algo:GenSam}. 
In the second stage, $r \ge r_0$ data points are sampled with replacement from the Big data through the optimal sub-sampling probabilities (estimated based on the model parameters obtained from stage one).
These two sub-sampled data sets are then combined to yield a single data set, and this data set is used to estimate the model parameters.  
Once these estimates have been obtained, $\tilde{\bm{V}}$, the variance-covariance matrix of $\tilde{\bm{\theta}}$, can be estimated via $\tilde{\bm{V}}=\tilde{\bm{J}}^{-1}_{\bm{X}} \tilde{\bm{V}}_c \tilde{\bm{J}}^{-1}_{\bm{X}}$, where
\begin{align}\label{Eq:V_theta}
        \tilde{\bm{J_X}} = & \frac{1}{N(r_0+r)} \Bigg[ \sum_{k \in \{r_0,r\}} \sum_{l=1}^{k} \frac{\ddot{u}(\tilde{\bm{\theta}}^T\bm{x}^k_l)  \bm{x}^k_l  (\bm{x}^k_l)^T [\ddot{\psi}(u(\tilde{\bm{\theta}}^T\bm{x}^k_l)) - y^k_l] + \ddot{\psi}(u(\tilde{\bm{\theta}}^T\bm{x}^k_l)) \dot{u}^2(\tilde{\bm{\theta}}^T\bm{x}^k_l) \bm{x}^k_l (\bm{x}^k_l)^T}{\phi^k_l} \Bigg] \notag \\
        \mbox{and} & \quad \tilde{\bm{V}_c} = \frac{1}{N^2(r_0+r)^2} \Bigg[\sum_{k \in \{r_0,r\}} \sum_{l=1}^{k} \frac{[y^k_l -\dot{\psi}(u(\tilde{\bm{\theta}}^T\bm{x}^k_l))]^2 \dot{u}^2(\tilde{\bm{\theta}}^T\bm{x}^k_l) \bm{x}^k_l (\bm{x}^k_l)^T}{ (\phi^k_l)^2} \Bigg].
\end{align}

Specific versions of the above algorithm under the logistic and Poisson regression models are given in Appendix~\ref{Appendix:Algorithms} in Algorithms~\ref{Algo:OSMAC} and \ref{Algo:OSPAC}, respectively.

\subsection{Limitations of the optimal sub-sampling algorithm}

The above optimal sub-sampling approach has a number of limitations.
One such limitation is the computational expense involved in obtaining the optimal sub-sampling probabilities, as these need to be found for each data point in the Big data set. 
To address this, \textcite{lee2021fast} introduced a faster two stage sub-sampling procedure for GLMs using the Johnson-Lindenstrauss Transform (JLT) and sub-sampled Randomised Hadamard Transform (SRHT), which are techniques to downsize matrix volume.
Another limitation is that, as the approximated optimal sub-sampling probabilities are proportional to $|y_i -\dot{\psi}(u(\tilde{\bm{\theta}}^T\bm{x}_i))|$, an observation with $y_i \approx \dot{\psi}(u(\tilde{\bm{\theta}}^T\bm{x}_i))$ has a near zero probability of being selected and data-points with $y_i = \dot{\psi}(u(\tilde{\bm{\theta}}^T\bm{x}_i))$ will be never sub-sampled.
\textcite{ai2021optimal} proposed to resolve this by introducing $\epsilon$ (a small positive value, e.g., $10^{-6}$) to constrain the optimal sub-sampling probabilities by replacing $[y_i -\dot{\psi}(u(\tilde{\bm{\theta}}^T\bm{x}_i))]$ with $\max{(|y_i -\dot{\psi}(u(\tilde{\bm{\theta}}^T\bm{x}_i))|,\epsilon)}$ which ensures such data points have a non-zero probability of being selected.
Lastly, one of the major limitations of the approach that has not been addressed previously is the inherent assumption that the Big data can be appropriately described by a given model. 
That is, the sub-sampling probabilities are evaluated based on an assumed model, and they are only optimal for this model.  
We suggest that this is a substantial limitation as specifying such a model in practice can be difficult.
This motivates the development of methods that yield sub-sampling probabilities that are robust to the choice of model, and our proposed approach for this is outlined next.

\section{Model robust optimal sub-sampling method}\label{Sec:ModRobOptSubMethod}

In order to apply the above two stage sub-sampling approach, optimal sub-sampling probabilities need to be evaluated, and these are based on a model that is assumed to appropriately describe the data. 
In practice, determining such a model may be difficult, and there could be a variety of models that appropriately describe the data.
Hence, a sub-sampling approach that provides robustness to the choice of model is desirable.  
For this, we propose to consider a set of $Q$ models which can be constructed to encapsulate a variety of scenarios that may be observed within the data.
For each model in this set, define model probabilities $\alpha_q$ for $q=1,\ldots,Q$ such that $\sum_{q=1}^Q\alpha_q=1$, which represents our {\it a priori} belief about the appropriateness of each model.
Denote the model matrix for the $q$-th model as $\bm{X}_q= h_q(\bm{X}_0)$, i.e., some function of the data matrix $\bm{X}_0$.
To apply a sub-sampling approach for this model set, sub-sampling probabilities are needed, and they should be constructed such that the resulting data is expected to address the analysis aim, regardless of which model is actually preferred for the Big data.  
For this purpose, we propose to form these sub-sampling probabilities by taking a weighted average (based on $\alpha_q)$ of the sub-sampling probabilities that would be obtained for each model (singularly).
This is the basic approach of our model robust optimal sub-sampling algorithm, for which further details are provided below, including a theoretical basis for constructing the sub-sampling probabilities in this way.

\subsection{Properties of model robust optimal sub-sampling algorithm}\label{Sec:ProforGLMModRob}

Updating the notation of the model matrix from $\bm{X}$ to $\bm{X}_q$ for the $q$-th model subsequently leads to analogous definitions for $\bm{x}_{qi}$, $\bm{\theta}_{q}$, $\hat{\bm{\theta}}_{qMLE}$, $\tilde{\bm{\theta}}_{q}$, ${\bm{J_X}}_q$, $\bm{V}_q$ and ${\bm{V}_q}_c$.
Assuming H\ref{Ass:1} to H\ref{Ass:7} hold for each of the $q$ models, Theorems~\ref{The:1} and \ref{The:2} apply straightforwardly to each of the models.
Extending the ideas from Section~\ref{Sec:Asymptotic}, optimal sub-sampling probabilities can be selected based on certain optimality criteria to, for example, ensure efficient estimates of parameters across the $Q$ models.  
This leads to the following theorems.

\begin{theorem}\label{The:5}
For a set of $Q$ models with model probability $\alpha_q$ for the $q$-th model, $q=1,\ldots Q$, if the sub-sampling probabilities are selected as follows:
    \begin{equation}
        \phi^{mMSE}_i = \sum_{q=1}^{Q} \alpha_q \frac{|y_i - \dot{\psi}(u(\hat{\bm{\theta}_q}_{MLE}^T\bm{x}_{qi}))|\, ||\bm{J}^{-1}_{\bm{X}_q} \dot{u}(\hat{\bm{\theta}_q}_{MLE}^T\bm{x}_{qi})\bm{x}_{qi} ||}{\sum_{j=1}^{N} |y_j - \dot{\psi}(u(\hat{\bm{\theta}_q}_{MLE}^T\bm{x}_{qj}))|\, ||\bm{J}^{-1}_{\bm{X}_q} \dot{u}(\hat{\bm{\theta}_q}_{MLE}^T\bm{x}_{qj})\bm{x}_{qj} ||},
    \end{equation}
    $i=1,\ldots,N$, and $\sum_{q=1}^{Q} \alpha_q = 1$, then $\sum_{q=1}^{Q} \alpha_q \mbox{tr}(\bm{V}_q)$ attains its minimum.
\end{theorem}

The proof of Theorem~\ref{The:5} is available in Appendix~\ref{Appendix:Theorem}, which is an extension of the proof of Theorem~\ref{The:3} in \textcite{ai2021optimal}.

\begin{theorem}\label{The:6}
    For a set of $Q$ models with model probability $\alpha_q$  for the $q$-th model, $q=1,\ldots Q$, if the sub-sampling probabilities are selected as follows: 
    \begin{equation}
        {\phi}_i^{mV_c} = \sum_{q=1}^{Q}\alpha_q \frac{|y_i - \dot{\psi}(u(\hat{\bm{\theta}_q}_{MLE}^T\bm{x}_{qi}))|\, || \dot{u}(\hat{\bm{\theta}_q}_{MLE}^T\bm{x}_{qi}) \bm{x}_{qi} ||}{\sum_{j=1}^{N} |y_j - \dot{\psi}(u(\hat{\bm{\theta}_q}_{MLE}^T\bm{x}_{qj}))|\, ||\dot{u}(\hat{\bm{\theta}_q}_{MLE}^T\bm{x}_{qj}) \bm{x}_{qj} ||},
    \end{equation}
    $i=1,\ldots,N$, and $\sum_{q=1}^{Q} \alpha_q = 1$, then $\sum_{q=1}^Q \alpha_q \mbox{tr}({\bm{V}_q}_c)$ attains its minimum.
\end{theorem}

The proof of Theorem \ref{The:6} follows straightforwardly from the proof of Theorem~\ref{The:5}.

The above theorems can be applied to obtain optimal sub-sampling probabilities under the logistic regression model as follows:
\begin{align}
    {\phi}^{mMSE}_i = \sum_{q=1}^{Q} \alpha_q \frac{|y_i - {\pi_q}_i| \, || \bm{J}^{-1}_{\bm{X}_q} \bm{x}_{qi} ||}{\sum_{j=1}^{N} |y_j - {\pi_q}_j| \, || \bm{J}^{-1}_{\bm{X}_q} \bm{x}_{qj} ||}, & \quad {\phi}^{mV_c}_i = \sum_{q=1}^{Q} \alpha_q \frac{|y_i - {\pi_q}_i| \, || \bm{x}_{qi}||}{\sum_{j=1}^{N} |y_j - {\pi_q}_j| \, || \bm{x}_{qj} ||}
\end{align}
with $y_i \in \{0,1\}$, ${\pi_q}_i = \exp{(\hat{\bm{\theta}}_{qMLE}^T\bm{x}_{qi})}/(1+\exp{(\hat{\bm{\theta}}_{qMLE}^T\bm{x}_{qi} )}),$  $\bm{J}_{\bm{X}_q}=N^{-1}\sum_{h=1}^{N} {\pi_q}_h(1-{\pi_q}_h) \bm{x}_{qh} (\bm{x}_{qh})^T$ and $i=1,\ldots,N$.

Similarly, optimal sub-sampling probabilities under the Poisson regression model can be obtained as follows:
\begin{align}
    {\phi}^{mMSE}_i = \sum_{q=1}^{Q} \alpha_q \frac{|y_i - {\lambda_q}_i|\,|| \bm{J}^{-1}_{\bm{X}_q} \bm{x}_{qi} ||}{\sum_{j=1}^{N} |y_j - {\lambda_q}_j|\,|| \bm{J}^{-1}_{\bm{X}_q} \bm{x}_{qj}||}, & \quad \phi^{mV_c}_i =  \sum_{q=1}^{Q} \alpha_q \frac{|y_i - {\lambda_q}_i|\,|| \bm{x}_{qi} || }{\sum_{j=1}^{N} |y_j - {\lambda_q}_j|\,|| \bm{x}_{qj} ||} 
\end{align}
with $y_i \in N_0$ or non-negative integers, ${\lambda_q}_i = \exp{(\hat{\bm{\theta}}_{qMLE}^T\bm{x}_{qi})},$ $\bm{J}_{\bm{X}_q}=N^{-1}\sum_{h=1}^{N} {\lambda_q}_h \bm{x}_{qh} (\bm{x}_{qh})^T$ and $i=1,\ldots,N$.

As noted in Section~\ref{Sec:Asymptotic} and by \textcite{ai2021optimal}, these model robust optimal sub-sampling probabilities are based on the maximum likelihood estimator found by considering the whole Big data set. 
Hence, a two stage procedure similar to Algorithm~\ref{Algo:OSGLMAC} is proposed for model robust sub-sampling, and this is outlined next.

\subsection{Model robust optimal sub-sampling algorithm for GLMs}\label{Sec:ModRobOptSubAlgforGLMs}

The two stage model robust optimal sub-sampling algorithm for GLMs is presented in Algorithm~\ref{Algo:OSGLMACMR}, where the estimates of sub-sampling probabilities based on the results of Theorems~\ref{The:5} and \ref{The:6} are used. 
The specific algorithms for logistic and Poisson regression are available in Appendix~\ref{Appendix:Algorithms} as Algorithms~\ref{Algo:OSMACMR} and \ref{Algo:OSPACMR}, respectively.
\newpage
\begin{algorithm}[H]
\SetAlgoLined

    \nonl \textbf{Stage 1} \\ \vspace{0.1cm}
     \textbf{Random Sampling :} Assign $\bm{\phi} = (\phi_1,\ldots,\phi_N)$ for $F_{N}$. For example, in logistic regression $\phi_i=\phi^{prop}$ represents proportional sub-sampling probabilities that are based on the response composition, and $\phi_i=N^{-1}$ which is uniform sub-sampling probabilities for Poisson regression.\\ \nonl 
    According to $\bm{\phi}$ draw random sub-samples of size $r_0$, such that  $S_{r_0}=\{h_q({\bm{x}_0}^{r_0}_{l}),y^{r_0}_l,\phi^{r_0}_l\}_{l=1}^{r_0} = (h_q(\bm{X}_0^{r_0}),\bm{y}^{r_0},\bm{\phi}^{r_0})$ for $q=1,\ldots,Q$. \\ 
    \textbf{Estimation:} For $q=1,\ldots,Q$ and $S_{r_0}$, find:
    \begin{equation*}
    \begin{aligned}
        \tilde{\bm{\theta}^{r_0}_{q}} = \argmaxA_{\bm{\theta}_q} ~  \log{L(\bm{\theta}_q| \bm{X}^{r_0}_{q},\bm{y}^{r_0},\bm{\phi}^{r_0})} \equiv \argmaxA_{\bm{\theta}_q} ~  \frac{1}{r_0}\sum_{l=1}^{r_0} \Big[\frac{y^{r_0}_l u(\bm{\theta}_q^T \bm{x}^{r_0}_{ql}) - \psi(u(\bm{\theta}_q^T\bm{x}^{r_0}_{ql})}{\phi^{r_0}_l} \Big].
    \end{aligned}
    \end{equation*} \nonl \\  
    \vspace{0.1cm} 
    \textbf{Stage 2} \\
    \vspace{0.1cm}
    \textbf{Optimal sub-sampling probability:} Estimate optimal sub-sampling probabilities $\bm{\phi}^{mMSE}$ or $\bm{\phi}^{mV_c}$ using the results from Theorem~\ref{The:5} or \ref{The:6}, with $\hat{\bm{\theta}}_{qMLE}$ replaced by $\tilde{\bm{\theta}}^{r_0}_{q}$.\\
    \textbf{Optimal sub-sampling and estimation:} Based on $\bm{\phi}^{mMSE}$ or $\bm{\phi}^{mV_c}$, draw sub-sample of size $r$ from $F_{N}$, such that  $S_{r}=\{h_q({\bm{x}_0}^r_{l}),\bm{y}^r_l,\bm{\phi}^r_l\}_{l=1}^{r} = (h_q(\bm{X}_0^{r}),\bm{y}^{r},\bm{\phi}^{r})$ for $q=1,\ldots,Q$. \\ \nonl 
    Combine $S_{r_0}$ and $S_{r}$ to form $S_{(r_0+r)}$, and obtain : 
    \begin{align*}
        \tilde{\bm{\theta}_{q}} = & \argmaxA_{\bm{\theta}_q} ~  \log{\big(L(\bm{\theta}_q| \bm{X}^{r_0}_q,\bm{y}^{r_0},\bm{\phi}^{r_0}, \bm{X}^{r}_q,\bm{y}^{r},\bm{\phi}^{r})\big)} \notag \\ 
        \equiv & \argmaxA_{\bm{\theta}_q} ~ \frac{1}{r_0+r} \Bigg[\sum_{k \in \{ r_0,r\}} \sum_{l=1}^{k} \frac{y^{k}_l u(\bm{\theta}^T_q\bm{x}^{k}_{ql}) - \psi(u(\bm{\theta}^T_q\bm{x}^{k}_{ql}))}{\phi^{k}_l} \Bigg].
    \end{align*} \\
    \textbf{Output:} $\tilde{\bm{\theta}}_q$ and $S_{(r_0+r)}$ for $q=1,\ldots,Q$.
\caption{Two stage model robust optimal sub-sampling algorithm for GLMs.}\label{Algo:OSGLMACMR}
\end{algorithm}

Similar to Algorithm~\ref{Algo:OSGLMAC}, the first stage entails randomly sub-sampling $F_{N}$ (with replacement), and estimating model parameters for each of the $Q$ models.
Based on these estimated parameters, model specific sub-sampling probabilities are obtained, and these are combined based on $\alpha_q$, forming model robust optimal sub-sampling probabilities. 
Subsequently, $r\ge r_0$ data points are sampled from $F_{N}$.
The two sub-samples are then combined, and each of the $Q$ models are fitted (separately) based on the weighted log-likelihood function, which should yield efficient estimates of parameters across each of the models. 
Similar to estimating the variance-covariance matrix of a single model (Equation \eqref{Eq:V_theta}), $\tilde{\bm{V}}_q$, the variance-covariance matrix of $\tilde{\bm{\theta}_q}$ can be estimated, for $q=1,\ldots,Q$. 

In the following section, the proposed model robust optimal sub-sampling method and the current optimal sub-sampling method are assessed through a simulation study and real world scenarios.

\section{Applications of optimal sub-sampling algorithms}\label{Sec:SimulationAndRealWorldSetup}

In this section, a simulation study and two real world applications are used to assess the performance of the proposed model robust optimal sub-sampling algorithm (Algorithm~\ref{Algo:OSGLMACMR}) compared to the optimal sub-sampling algorithm (Algorithm~\ref{Algo:OSGLMAC}), and random sampling.
The main results are presented in this section with some results presented in the Appendix.
The simulation study and real world applications were coded in the R statistical programming language with the help of Rstudio IDE, and our scripts are available through Github. 
Appendix~\ref{Appendix:Code} provides specific Github hyperlinks to the code repositories for the simulation study and real world applications.

\subsection{Simulation study design}\label{Sec:Simulation}

To explore the performance of our model robust sub-sampling approach, a simulation study was constructed based on the logistic and Poisson regression models.
For each case, a set of $Q=4$ models were assumed based on \textcite{shi2021model}, and this set is summarised in Table~\ref{Tab:1}.
For each model, $F_{N}$ was constructed, by assuming a distribution for the covariates and corresponding response.
The performance of the three sampling methods were then compared for each $F_{N}$ through evaluating six scenarios: 1) random sampling to estimate the parameters of the data generating model; 2) optimal sub-sampling under the data generating model -- 
this simulates the case where an appropriate model was assumed for describing the Big data; 3)-5) optimal sub-sampling under alternative models i.e., optimal sub-sampling to estimate the parameters of the data generating model but where samples were obtained based on an alternative model -- this simulates undertaking optimal sub-sampling based on a `wrong' model; 6) model robust optimal sub-sampling, with $\alpha_q=1/4$ used to estimate the parameters of the data generating model, for $q=1,\ldots,4$.
To quantitatively compare each approach, the simulated Mean Squared Error (SMSE) was used. This was evaluated as follows:
\begin{equation}\label{Eq:SMSE}
    SMSE(\bm{\theta}) = \frac{1}{M}{\sum_{m=1}^{M} \sum_{n=1}^{p+q} (\tilde{\theta}_{nm} - \theta_n)^2 },
\end{equation}
where $M$ is the number of simulations, $p+q$ is the number of parameters, $\theta_n$ is the $n$-th underlying parameter of the data generating model and $\tilde{\theta}_{nm}$ is the estimate of this parameter from the $m$-th simulation.

In addition, the determinant of the observed Fisher information matrix (i.e., the inverse of the variance-covariance matrix $\bm{V}$) was also compared across sub-sampling approaches, with the largest of such values being preferred.  
For each simulation, $N=10000$, $r_0=100$, $r=100,200,\ldots,1400$ and $M=1000$.
\begin{table}[H]
    \centering
    \caption{Model set assumed for simulation study.}\label{Tab:1}
    \begin{tabular}{l} \hline
        \textbf{Model set}  \\ \hline 
         $\theta_0 + \theta_1 \bm{x}_1 + \theta_2 \bm{x}_2$ \\ 
         $\theta_0 + \theta_1 \bm{x}_1 + \theta_2 \bm{x}_2 + \theta_3 \bm{x}^2_1$  \\ $\theta_0 + \theta_1 \bm{x}_1 + \theta_2 \bm{x}_2 + \theta_3 \bm{x}^2_2$  \\ $\theta_0 + \theta_1 \bm{x}_1 + \theta_2 \bm{x}_2 + \theta_3 \bm{x}^2_1 + \theta_4 \bm{x}^2_2$ \\ \hline
    \end{tabular}
\end{table}

Additionally, we only consider the $A$-optimal ($mMSE$) sub-sampling strategy in these simulations, as \textcite{wang2018logistic,ai2021optimal} indicated that this approach generally outperformed the $L$-optimal ($mV_c$) strategy.
We explore both optimality criteria in the real-world applications.

\subsubsection{Logistic regression}\label{Sec:LogisticRegression}

Following \textcite{wang2018logistic}, covariate data for the logistic regression model were simulated based on two distributions: Exponential ($\lambda$) and Multivariate Normal ($\bm{\mu},\bm{\Sigma}$). 
The values of ($\lambda,\bm{\mu},\bm{\Sigma}$) and $\bm{\theta}$ are given in Table~\ref{Tab:3} for each data generating model. 
For all models, the first element of $\bm{\theta}$ is the value of the parameter for the intercept.
While $\lambda$, $\bm{\mu}$ and $\bm{\Sigma}$ were selected arbitrarily, $\bm{\theta}$ was selected such that for $N=10000$, the data generating model was preferred over the alternative models based on the Akaike Information Criterion \parencite{akaike1974new}.
\begin{table}[H]
    \centering
    \caption{$\lambda,\bm{\mu},\bm{\Sigma}$ and $\bm{\theta}$ values are given to generate $x_1,x_2$  through Exponential and Multivariate Normal distributions and form $F_{N}$ for each data generating logistic regression model.}\label{Tab:3}
    \begin{tabular}{lll} \hline
        \multirow{3}{*}{\textbf{Covariates}} & \multicolumn{2}{c}{\textbf{Distributions}} \\ \cline{2-3} 
        & \textbf{Exponential} ($\lambda=\sqrt{3}$) & \textbf{Normal} $\Big(\bm{\mu}=[0,0]$,$\bm{\Sigma}= \begin{bmatrix} 1.5 & 0 \\ 0 & 1.5 \end{bmatrix}\Big)$ \\ \hline
        $\bm{x}_1,\bm{x}_2$ &  $\bm{\theta}=[-2,\hphantom{-}1.5,\hphantom{-}0.3]$ & $\bm{\theta}=[-1,\hphantom{-}0.5,\hphantom{-}0.1]$\\ 
        $\bm{x}_1, \bm{x}_2, \bm{x}^2_1$ &  $\bm{\theta}=[-2,\hphantom{-}1.7,-1.2,\hphantom{-}0.2]$ & $\bm{\theta}=[-1,\hphantom{-}0.5,\hphantom{-}0.5,\hphantom{-}0.7]$ \\ 
        $\bm{x}_1, \bm{x}_2, \bm{x}^2_2$ &  $\bm{\theta}=[-2,-1.3,\hphantom{-}1.9,\hphantom{-}0.9]$ & $\bm{\theta}=[-1,\hphantom{-}0.5,\hphantom{-}0.5,\hphantom{-}0.3]$ \\ 
        $\bm{x}_1, \bm{x}_2, \bm{x}^2_1, \bm{x}^2_2$ &  $\bm{\theta}=[-2,\hphantom{-}1.9,\hphantom{-}1.9,\hphantom{-}0.9,\hphantom{-}0.7]$ & $\bm{\theta}=[-1,\hphantom{-}0.5,\hphantom{-}0.5,\hphantom{-}0.5,\hphantom{-}0.5]$ \\ \hline
    \end{tabular}
\end{table}

\begin{figure}[H]
    \centering
    \includegraphics[width=\textwidth,height=8.25in]{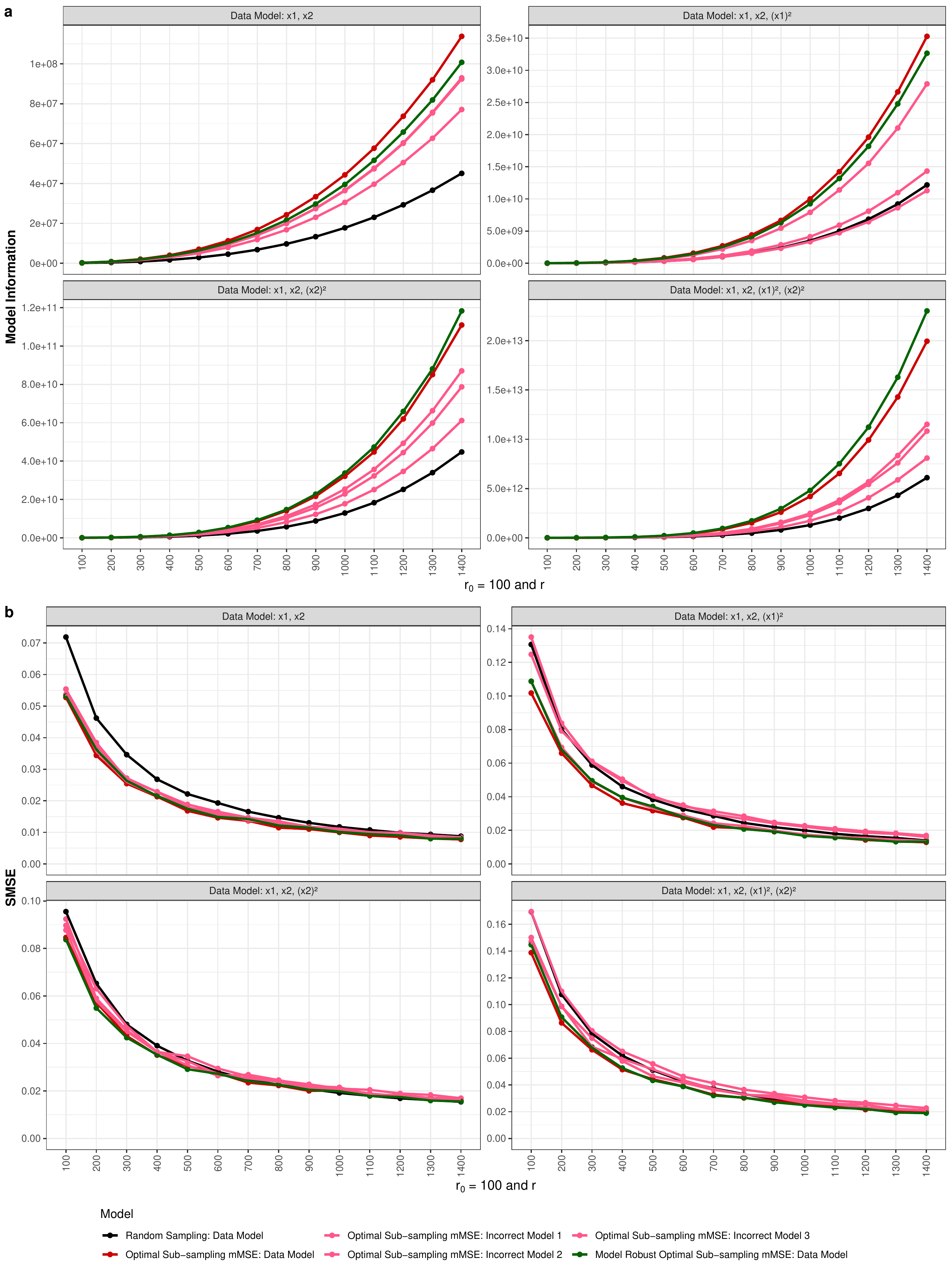}
    \caption{(a) Model information (larger is better) and (b) SMSE (smaller is better) for the sub-sampling methods for the logistic regression model under $mMSE$. Covariate data were generated from the Multivariate Normal distribution.} \label{Fig:RWA_LR_TV_Nor}
\end{figure}

Figure~\ref{Fig:RWA_LR_TV_Nor} provides summaries of the SMSE and average model information over all models when the covariate data is generated through a Multivariate Normal distribution.
The SMSE and average model information indicate that, under optimal sub-sampling for $mMSE$, the data generating model is typically preferred within the model set. 
This is expected as it is the case where the appropriate data generating model was correctly assumed to describe the Big data.
Of note, the proposed model robust approach performs similarly to the optimal sub-sampling approach.
Notable increases in the SMSE and decreases in the model information are observed when the incorrect model is considered for optimal sub-sampling compared to the data generating model.
Overall, random sampling tends to have the worst performance.
Similar results were obtained when the covariate data were generated through an Exponential distribution (Appendix~\ref{Appendix:Figures}).

\subsubsection{Poisson regression}\label{Sec:PoissonRegression}
	
The simulation study based on Poisson regression was constructed similar to the logistic regression case.  
In terms of generating covariate values, Uniform and Multivariate Normal ($\bm{\mu}, \bm{\Sigma}$) distributions were used, see Table~\ref{Tab:5} and \textcite{ai2021optimal}.
Values for $\bm{\theta}$ were selected as described above, and are given in this table.
\begin{table}[htbp!]
    \centering
    \caption{$\bm{\mu},\bm{\Sigma}$ and $\bm{\theta}$ values are given to generate $x_1,x_2$ through uniform and Multivariate Normal distributions and form $F_{N}$ for each data generating Poisson regression model.}
    \label{Tab:5}
    \begin{tabular}{lll} \hline
        \multirow{3}{*}{\textbf{Covariates}} & \multicolumn{2}{c}{\textbf{Distributions}} \\ \cline{2-3} 
        & \textbf{Uniform} & \textbf{Normal} $\Big(\bm{\mu}=[0,0]$,$\bm{\Sigma}= \begin{bmatrix} 1 & 0 \\ 0 & 1 \end{bmatrix}\Big)$ \\ \hline
        $\bm{x}_1,\bm{x}_2$ & $\bm{\theta}=[1,\hphantom{-}0.3,\hphantom{-}0.3]$ & $\bm{\theta}=[1,\hphantom{-}0.5,\hphantom{-}0.1]$\\ 
        $\bm{x}_1, \bm{x}_2, \bm{x}^2_1$ & $\bm{\theta}=[1,\hphantom{-}0.7,-0.5,\hphantom{-}0.4]$ & $\bm{\theta}=[1,-0.3,\hphantom{-}0.3,\hphantom{-}0.1]$ \\ 
        $\bm{x}_1, \bm{x}_2, \bm{x}^2_2$ & $\bm{\theta}=[1,-0.4,\hphantom{-}0.5,\hphantom{-}0.3]$ & $\bm{\theta}=[1,\hphantom{-}0.3,-0.3,\hphantom{-}0.1]$ \\ 
        $\bm{x}_1, \bm{x}_2, \bm{x}^2_1, \bm{x}^2_2$ & $\bm{\theta}=[1,\hphantom{-}0.5,\hphantom{-}0.5,-0.3,\hphantom{-}0.5]$ & $\bm{\theta}=[1,\hphantom{-}0.3,\hphantom{-}0.3,-0.3,\hphantom{-}0.1]$ \\ \hline
    \end{tabular}
\end{table}	
	
\begin{figure}[H]
    \centering
    \includegraphics[width=\textwidth,height=8.25in]{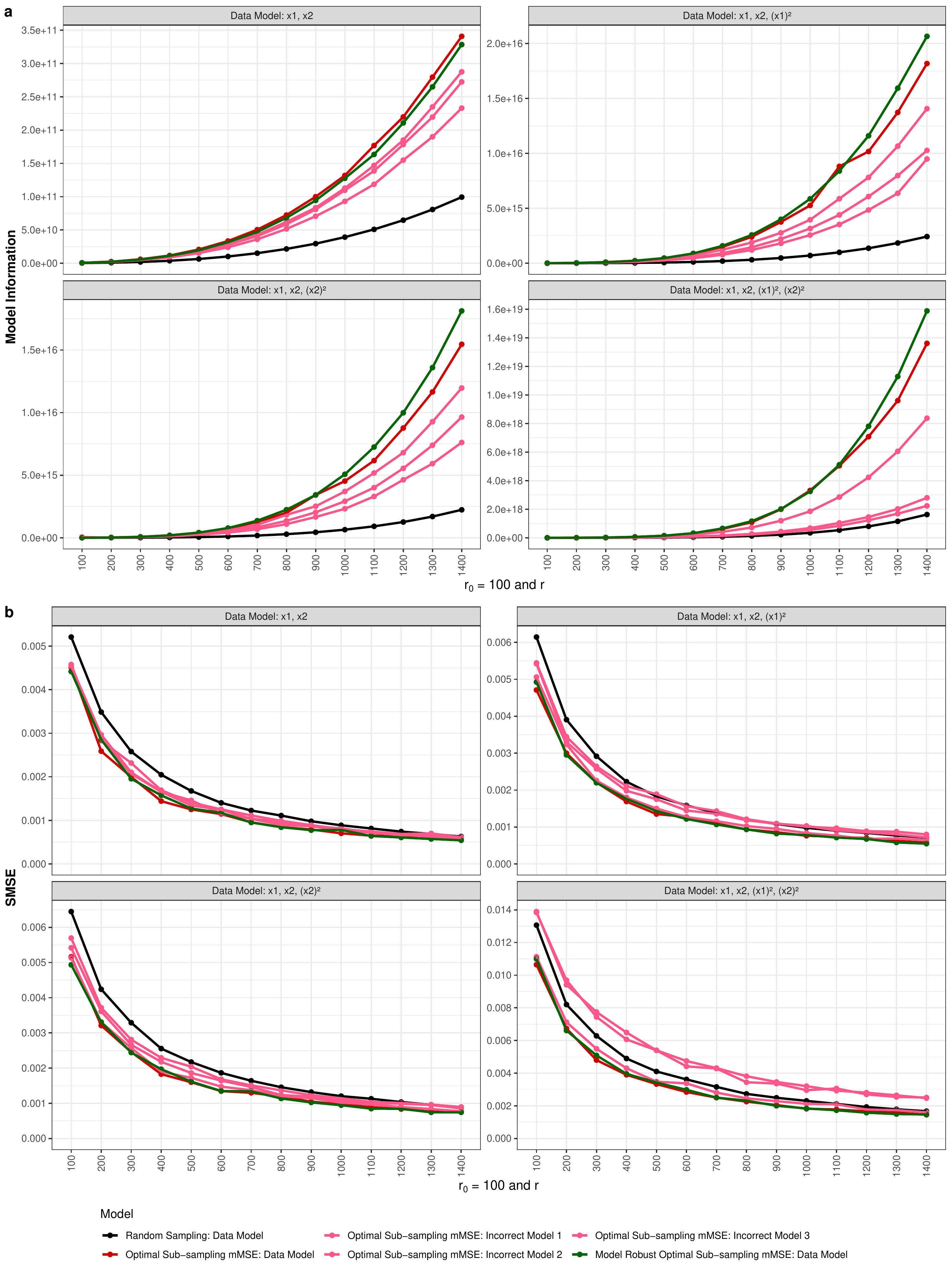} 
    \caption{(a) Model information (larger is better) and (b) SMSE (smaller is better) for the sub-sampling methods for the Poisson regression model under $mMSE$. Covariate data were generated from the standardised Multivariate Normal distribution.}
    \label{Fig:RWA_PR_TV_Nor}
\end{figure}

SMSE and average model information over all models when the covariate data was generated from a Multivariate Normal distribution and the response from generated form a Poisson regression model are shown in Figure~\ref{Fig:RWA_PR_TV_Nor}. 
Generally, random sampling performs worst, while the proposed model robust approach and the optimal sub-sampling method based on the data generating model perform the best, and have similar SMSE and average model information values.  
Again, the use of the optimal sub-sampling algorithm can lead to notable increases in SMSE when the assumed model is incorrect.

Similar results were obtained when the covariate data was generated from a uniform distribution (see Appendix~\ref{Appendix:Figures}).

\subsection{Real world applications}\label{Sec:RealWorldApplications}

The three sub-sampling methods are applied to analyse the ``Skin segmentation" and ``New York City taxi fare" data under logistic and Poisson regression, respectively.
In the simulation study, the parameters were specified for the data generating model. 
However, in real world applications these are unknown.
In this situation the sub-sampling methods cannot be compared as in Section \ref{Sec:Simulation}.
Instead, for every model, the simulated mean squared error (SMSE) under each sub-sampling method is evaluated for various $r$ sub-sample sizes and $M$ simulations.
This can be evaluated as follows:
\begin{equation}
    SSMSE(\hat{\bm{\theta}}_{MLE}) = \sum_{q=1}^{Q} SMSE_q(\hat{\bm{\theta}}_{MLE}),
\end{equation}
where $SMSE_q(\hat{\bm{\theta}}_{MLE})$ is the simulated mean squared error in Equation \eqref{Eq:SMSE} with $\bm{\theta}$ replaced by  $\hat{\bm{\theta}}_{MLE}$. 

In the following real world examples, the set of $Q$ models includes the main effects model, with intercept, and all possible combinations of quadratic terms for continuous covariates.  
Again, these were constructed based on the work of \textcite{shi2021model}.
	
\subsubsection{Identifying skin from colours in images} \label{Sec:RealLogisticRegression}

\textcite{Skin_Data} considered the problem of identifying skin-like regions in images as part of the complex task of performing facial recognition.
For this purpose, \textcite{Skin_Data} collated the ``Skin segmentation" data set, which consists of RGB (R-red, G-green, B-blue) values of randomly sampled pixels from $N=245,057$ face images (out of which $50,859$ are skin samples and $194,198$ are non-skin samples) from various age groups, race groups and genders. 
\textcite{bhatt2009efficient,binias2018pixel} applied multiple supervised machine learning algorithms to classify if images are skin or not based on the RGB colour data.
In addition, \textcite{abbas2019skin} conducted the same classification task for two different colour spaces, HSV (H-hue, S-saturation, V-value) and YCbCr (Y-luma component, Cb-blue difference chroma component, Cr-red difference chroma component), by transforming the RGB colour space.
	
We consider the same classification problem but use a logistic regression model. 
Skin presence is denoted as one and skin absence is denoted as zero. 
Each colour vector is scaled to have a mean of zero and a variance of one (initial range was between $0-255$).
To compare sub-sampling methods, we set $r_0=200,r=200,300,\ldots,1800$ for the sub-samples, and construct a set of $Q$ models by considering an intercept with main effects model with all covariates (scaled colors red,green and blue) as the base model, and form all alternative models by including different combinations of quadratic terms of all covariates.  
This leads to $Q=8$.
Each of these models is considered equally likely {\it a priori}.

\begin{figure}[H]
    \centering
    \includegraphics[width=\textwidth,height=4in]{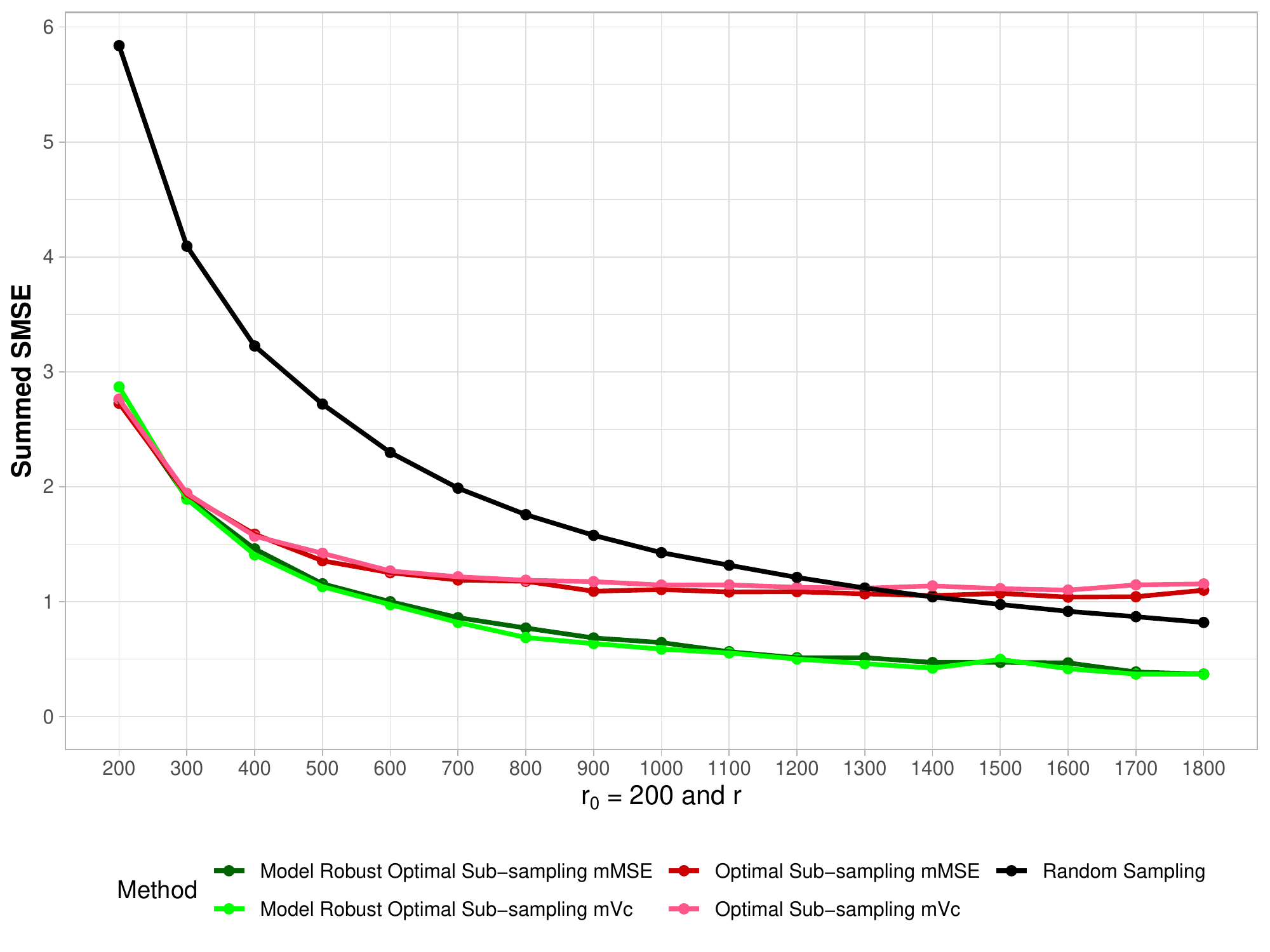}
    \caption{Summed SMSE over the available models for logistic regression applied on the ``Skin segmentation" data.}
    \label{Fig:RWS_SkinData}
\end{figure}

Figure~\ref{Fig:RWS_SkinData} shows the SSMSE values over the $Q$ models for various sub-sample sizes obtained by applying logistic regression to the ``Skin segmentation" data.
The proposed model robust approach performs similarly to the optimal sub-sampling method for $r=200$ and $r=300$.  
However, when the sample size increases the optimal sub-sampling method performs poorly, and after $r=1300$ random sampling actually has lower SSMSE values than optimal sub-sampling.  
The same is not true for our proposed model robust approach which has the lowest values of SSMSE throughout the selected sample sizes under $mMSE$ and $mV_c$ optimal sub-sampling criteria.
It is interesting that the optimal sub-sampling approach based on $mMSE$ performed worse than random sampling in comparison to the simulation study results.
Upon investigating this, it was found that it could potentially be explained by one of the models being particularly poor (subsequent higher SMSE with increasing $r$ values) for describing the data, and therefore led to inflated SSMSE values. 
Despite this, we note that the model robust approach appears to perform well in general.

\subsubsection{New York City taxi cab usage}\label{Sec:RealPoissonRegression}

New York City (NYC) taxi trip and fare information from 2009 onward, consisting of over $170$ million records each year, are publicly available courtesy of the New York City Taxi and Limousine Commission.
Some analyses of interest of these NYC taxi data include: a taxi driver's decision process to pick up a fair or cruise for customers which was modelled via a logistic regression model \parencite{yazici2013big}; taxi demand and how it is impacted by location, time, demographic information, socioeconomic status and employment status which was modelled via a multiple linear regression model \parencite{yang2014modeling}; and the dependence of taxi supply and demand on location and time considered via the Poisson regression model \parencite{yang2017modeling}.

In our application, we are interested in how taxi usage varies with day of the week (weekday/weekend), season (winter/spring/summer/autumn), fare amount, and cash payment. 
The data used in our application is the ``New York City taxi fare'' data for the year 2013, hosted by the University of Illinois Urbana Champaign \parencite{illinoisdatabankIDB}. 

Each data point includes the number of rides recorded against the medallion number (a license number provided for a motor vehicle to operate as a taxi) ($y$), weekday or not ($\bm{x_1}$), winter or not ($\bm{x_2}$), spring or not ($\bm{x_3}$), summer or not ($\bm{x_4}$), summed fare amount in dollars ($\bm{x_5}$), and the ratio of cash payment trips in terms of all trips ($\bm{x_6}$).
The continuous covariate $\bm{x_5}$ was scaled to have a range of zero and one.
Poisson regression was used to model the relationship between the number of rides per medallion and these covariates.

\begin{figure}[H]
    \centering
    \includegraphics[width=\textwidth,height=4in]{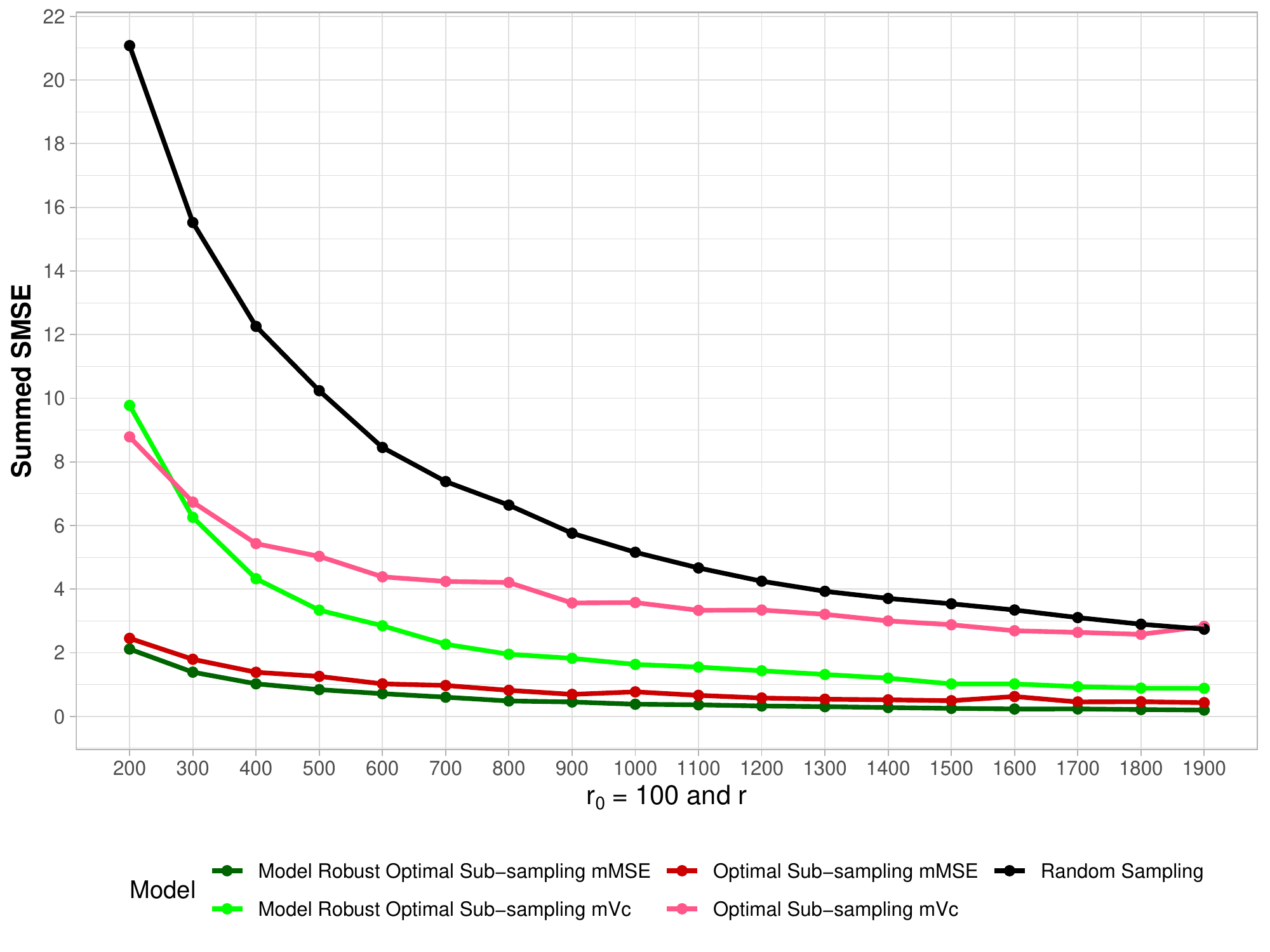}
    \caption{Summed SMSE over the available models for Poisson regression applied on the 2013 NYC taxi usage data.}
    \label{Fig:RWS_NYCTaxiFareData}
\end{figure}

For our study, three sub-sampling methods are compared for the analysis of the taxi fare data, assigning $r_0=100$ and $r=200,\ldots,1900$ for sub-samples. 
The model set consists of the main effects model($\bm{x_1},\bm{x_2},\bm{x_3},\bm{x_4},\bm{x_5},\bm{x_6}$) and all possible combinations of the quadratic terms of the continuous covariates ($\bm{x^2_5},\bm{x^2_6}$) which leads to $Q=4$. 
Each of these models were considered equally likely {\it a priori}.

The SSMSE over the four models is shown in Figure~\ref{Fig:RWS_NYCTaxiFareData}.
Our proposed model robust approach outperforms the optimal sub-sampling method for almost all sample sizes for both $mMSE$ and $mV_c$ strategy. 
Under $mV_c$,  the model robust approach actually initially performs worse than the optimal sub-sampling approach but this is quickly reserved as $r$ is increased.
Random sampling performs the worst, suggesting that there is benefit in using targeted sampling approaches (as proposed here) over random selection.

\section{Discussion}\label{Sec:Discussion}

In this article, we proposed a model robust optimal sub-sampling approach for GLMs.
This new approach extends the current optimal sub-sampling approach by considering a set of models (rather than a single model) when determining the sub-sampling probabilities.  
The exact formulation of these probabilities is derived using Theorems~\ref{The:5} and \ref{The:6}.
The robustness properties of this proposed approach were demonstrated in a simulation study, and in two real-world analysis problems, where it was shown to outperform optimal sub-sampling and random sampling.
Accordingly, we suggest that such an approach be considered in future Big data analysis problems.

The main limitation of the proposed approach that could be addressed in future research is extending the specification of the model set to a flexible class of models.  
This could be, for example, through a generalised additive model \parencite{hastie1986generalized} or the inclusion of a discrepancy term in the linear predictor \parencite{krishna2021robust}.  
Another avenue of interest that could also be explored is reducing the model set after stage one of the two-stage algorithm where models that clearly do not appear to be appropriate for the data could be dropped.
Both of these extensions are planned for future research.

\printbibliography[title={References},heading=bibnumbered]

@article{abbas2019skin,
  title={{Skin Detection using Improved ID3 Algorithm}},
  author={Abbas, Ayad R and Farooq, Ayat O},
  journal={Iraqi Journal of Science},
  volume={60},
  number={2}, 
  pages={402--410},
  year={2019},
  url={https://ijs.uobaghdad.edu.iq/index.php/eijs/article/view/658}
}

@article{ai2020quantile,
  title={{Optimal subsampling for large-scale quantile regression}},
  author={Ai, Mingyao and Wang, Fei and Yu, Jun and Zhang, Huiming},
  journal={Journal of Complexity},
  volume = {62},
  pages = {101512},
  year = {2021},
  issn = {0885-064X},
  publisher={Elsevier},
  doi={https://doi.org/10.1016/j.jco.2020.101512}
}

@article{ai2021optimal,
  title={{Optimal subsampling algorithms for Big data regressions}},
  author={Ai, Mingyao and Yu, Jun and Zhang, Huiming and Wang, HaiYing},
  journal={Statistica Sinica},
  volume={31},
  pages={749--772},
  year={2021},
  doi={https://doi.org/10.5705/ss.202018.0439}
}

@article{akaike1974new,
  title={{A new look at the statistical model identification}},
  author={Akaike, Hirotugu},
  journal={IEEE Transactions on Automatic Control},
  volume={19},
  number={6},
  pages={716--723},
  year={1974},
  publisher={IEEE},
  doi={https://doi.org/10.1109/TAC.1974.1100705}
}

@book{atkinson2007optimum,
  title={{Optimum Experimental Designs, with SAS}},
  author={Atkinson, Anthony and Donev, Alexander and Tobias, Randall},
  isbn={9780191537943},
  lccn={2007299129},
  series={Oxford Statistical Science Series},
  year={2007},
  publisher={OUP Oxford}
}

@inproceedings{bhatt2009efficient,
  title={{Efficient Skin Region Segmentation Using Low Complexity Fuzzy Decision Tree Model}},
  author={Bhatt, Rajen B and Sharma, Gaurav and Dhall, Abhinav and Chaudhury, Santanu},
  booktitle={2009 Annual IEEE India Conference},
  pages={1--4},
  year={2009},
  organization={IEEE},
  doi={https://doi.org/10.1109/INDCON.2009.5409447}
}

@incollection{binias2018pixel,
  title={{Pixel Classification for Skin Detection in Color Images}},
  author={Binias, Bartosz and Fr{\k a}ckiewicz, Mariusz and Jaskot, Krzysztof and Palus, Henryk},
  booktitle={Advanced Technologies in Practical Applications for National Security},
  pages={87--99},
  year={2018},
  isbn={978-3-319-64674-9},
  publisher={Springer International Publishing},
  doi={https://doi.org/10.1007/978-3-319-64674-9_6}
}

@article{chang2017divide,
  title={{Divide and conquer local average regression}},
  author={Xiangyu Chang and Shao-Bo Lin and Yao Wang},
  journal={Electronic Journal of Statistics},
  volume={11},
  number={1},
  pages={1326--1350},
  year={2017},
  publisher={The Institute of Mathematical Statistics and the Bernoulli Society},
  doi={https://doi.org/10.1214/17-EJS1265}
}

@article{cheng2020IBOSSlogistic,
  title={{Information-based optimal subdata selection for big data logistic regression}},
  author={Cheng, Qianshun and Wang, Haiying and Yang, Min},
  journal={Journal of Statistical Planning and Inference},
  year={2020},
  volume = {209},
  pages = {112-122},
  issn = {0378-3758},
  publisher={Elsevier},
  doi={https://doi.org/10.1016/j.jspi.2020.03.004}
}

@inproceedings{clemenccon2014scaling,
  title={{Scaling up M-estimation via sampling designs: The Horvitz-Thompson stochastic gradient descent}},
  author={Cl{\'e}men{\c{c}}on, St{\'e}phan and Bertail, Patrice and Chautru, Emilie},
  booktitle={2014 IEEE International Conference on Big Data (Big Data)},
  pages={25--30},
  year={2014},
  organization={IEEE},
  doi={https://doi.org/10.1109/BigData.2014.7004208}
}

@article{cleveland2014divide,
  title={{Divide and recombine (D\& R): Data science for large complex data}},
  author={Cleveland, William, S and Hafen, Ryan},
  journal={Statistical Analysis and Data Mining: The ASA Data Science Journal},
  volume={7},
  number={6},
  pages={425--433},
  year={2014},
  publisher={Wiley Online Library},
  doi={https://doi.org/10.1002/sam.11242}
}

@misc{illinoisdatabankIDB,
  author = {Donovan, Brian and Work, Dan},
  publisher = {University of Illinois at Urbana-Champaign},
  title = {{New York City Taxi Trip Data (2010-2013)}},
  year = {2016},
  doi = {https://doi.org/10.13012/J8PN93H8}
}

@article{drovandi2017principles,
  title={{Principles of Experimental Design for Big Data Analysis}},
  author={Drovandi, Christopher C and Holmes, Christopher and McGree, James M and Mengersen, Kerrie and Richardson, Sylvia and Ryan, Elizabeth G},
  journal={Statistical Science},
  volume={32},
  number={3},
  pages={385--404},
  year={2017},
  publisher = {Institute of Mathematical Statistics},
  doi={https://doi.org/10.1214/16-STS604}
}

@article{fahrmeir1985consistency,
  title={{Consistency and Asymptotic Normality of the Maximum Likelihood Estimator in Generalized Linear Models}},
  author={Fahrmeir, Ludwig and Kaufmann, Heinz},
  journal={The Annals of Statistics},
  volume={13},
  number={1},
  pages={342--368},
  year={1985},
  publisher={Institute of Mathematical Statistics},
  doi={https://doi.org/10.1214/aos/1176346597}
}

@article{guha2012large,
  title={{Large complex data: divide and recombine (D \& R) with RHIPE}},
  author={Guha, Saptarshi and Hafen, Ryan and Rounds, Jeremiah and Xia, Jin and Li, Jianfu and Xi, Bowei and Cleveland, William, S},
  journal={Stat},
  volume={1},
  number={1},
  pages={53--67},
  year={2012},
  publisher={Wiley Online Library},
  doi={https://doi.org/10.1002/sta4.7}
}

@article{hastie1986generalized,
  title={{Generalized Additive Models}},
  author={Hastie, Trevor and Tibshirani, Robert},
  journal={Statistical Science},
  publisher = {Institute of Mathematical Statistics},
  volume = {1},
  number = {3},
  pages={297--310},
  year={1986},
  ISSN = {08834237},
  url={https://www.jstor.org/stable/2245459}
}

@article{karmakar2018statistical,
  title={{Statistical Validity and Consistency of Big Data Analytics: A General Framework}},
  author={Karmakar, Bikram and Mukhopadhyay, Indranil},
  volume={18},
  number={2},
  year={2020},
  journal={Statistics and Applications},
  pages={369 -- 381},
  url={https://ssca.org.in/media/25_Vol._18_No._2_2020_SA_Indranil_Mukhopadhyay.pdf}
}

@article{kleiner2014scalable,
  title={{A scalable bootstrap for massive data}},
  author={Kleiner, Ariel and Talwalkar, Ameet and Sarkar, Purnamrita and Jordan, Michael, I},
  journal={Journal of the Royal Statistical Society: Series B: Statistical Methodology},
  pages={795--816},
  year={2014},
  volume = {76},
  number = {4},
  publisher={JSTOR},
  doi={https://doi.org/10.1111/rssb.12050}
}

@article{krishna2021robust,
  title={{Robust experimental designs for model calibration}},
  volume = {0},
  number = {0},
  author={Krishna, Arvind and Joseph, V Roshan and Ba, Shan and Brenneman, William A and Myers, William R},
  journal={Journal of Quality Technology},
  pages={1--12},
  year={2021},
  publisher={Taylor \& Francis},
  doi={https://doi.org/10.1080/00224065.2021.1930618}
}

@article{lee2021fast,
  title={{Fast Optimal Subsampling Probability Approximation for Generalized Linear Models}},
  author={Lee, JooChul and Schifano, Elizabeth D and Wang, HaiYing},
  journal={Econometrics and Statistics},
  year={2021},
  issn = {2452-3062},
  publisher={Elsevier},
  doi={https://doi.org/10.1016/j.ecosta.2021.02.007}
}

@article{li2020sequential,
  title={{A sequential split-and-conquer approach for the analysis of big dependent data in computer experiments}},
  author={Li, Chengrui and Hung, Ying and Xie, Minge},
  journal={Canadian Journal of Statistics},
  volume={48},
  number={4},
  pages={712--730},
  year={2020},
  publisher={Wiley Online Library},
  doi={https://doi.org/10.1002/cjs.11559}
}

@article{lin2011aggregated,
  title={{Aggregated estimating equation estimation}},
  author={Lin, Nan and Xi, Ruibin},
  journal={Statistics and Its Interface},
  volume={4},
  number={1},
  pages={73--83},
  year={2011},
  publisher={International Press of Boston},
  doi={https://doi.org/10.4310/SII.2011.v4.n1.a8}
}

@article{ma2015statistical,
  title={{A Statistical Perspective on Algorithmic Leveraging}},
  author={Ma, Ping and Mahoney, Michael, W and Yu, Bin},
  journal={Journal of Machine Learning Research},
  volume={16},
  number = {27},
  pages={861--911},
  year={2015},
  publisher={JMLR. org},
  url = {http://jmlr.org/papers/v16/ma15a.html}
}

@article{ma2015leveraging,
  title={{Leveraging for big data regression}},
  author={Ma, Ping and Sun, Xiaoxiao},
  journal={Wiley Interdisciplinary Reviews: Computational Statistics},
  volume={7},
  number={1},
  pages={70--76},
  year={2015},
  publisher={Wiley Online Library},
  doi={https://doi.org/10.1002/wics.1324}
}

@article{martinez2020modeling,
  title={{Modeling and Management Big Data in Databases—A Systematic Literature Review}},
  author={Martinez-Mosquera, Diana and Navarrete, Rosa and Lujan-Mora, Sergio},
  journal={Sustainability},
  volume={12},
  number={2},
  pages={634},
  year={2020},
  ISSN = {2071-1050},
  publisher={Multidisciplinary Digital Publishing Institute},
  doi={https://doi.org/10.3390/su12020634}
}

@article{meng2021lowcon,
  title={{LowCon: A Design-based Subsampling Approach in a Misspecified Linear Model}},
  author={Meng, Cheng and Xie, Rui and Mandal, Abhyuday and Zhang, Xinlian and Zhong, Wenxuan and Ma, Ping},
  journal={Journal of Computational and Graphical Statistics},
  volume={30},
  number={3},
  pages={694--708},
  year={2021},
  publisher={Taylor \& Francis},
  doi={https://doi.org/10.1080/10618600.2020.1844215}
}

@article{nelder1972generalized,
  title={{Generalized Linear Models}},
  author={Nelder, John Ashworth and Wedderburn, Robert WM},
  journal={Journal of the Royal Statistical Society: Series A (General)},
  volume={135},
  number={3},
  pages={370--384},
  year={1972},
  publisher={Wiley Online Library},
  doi={https://doi.org/10.2307/2344614}
}

@article{nguyen2020systematic,
  title={{A Systematic Review of Big Data Analytics for Oil and Gas Industry 4.0}},
  author={Nguyen, Trung and Gosine, Raymond G and Warrian, Peter},
  journal={IEEE access},
  volume={8},
  pages={61183--61201},
  year={2020},
  publisher={IEEE},
  doi={https://doi.org/10.1109/ACCESS.2020.2979678}
}

@misc{Skin_Data ,
 title={{Skin Segmentation}},
 author = {Rajen, Bhatt and Abhinav, Dhall},
 year = {2012},
 howpublished = {UCI Machine Learning Repository},
 url = {https://archive.ics.uci.edu/ml/datasets/skin+segmentation}
}

@article{rehman2021leveraging,
  title={{Leveraging big data analytics in healthcare enhancement: trends, challenges and opportunities}},
  author={Rehman, Arshia and Naz, Saeeda and Razzak, Imran},
  journal={Multimedia Systems},
  pages={1--33},
  year={2021},
  publisher={Springer},
  doi={https://doi.org/10.1007/s00530-020-00736-8}
}

@article{schifano2016online,
  title={{Online Updating of Statistical Inference in the Big Data Setting}},
  author={Schifano, Elizabeth, D and Wu, Jing and Wang, Chun and Yan, Jun and Chen, Ming-Hui},
  journal={Technometrics},
  volume={58},
  number={3},
  pages={393--403},
  year={2016},
  publisher={Taylor \& Francis},
  doi={https://doi.org/10.1080/00401706.2016.1142900}
}

@article{schwarz1978estimating,
  title={{Estimating the Dimension of a Model}},
  author={Schwarz, Gideon},
  journal={The Annals of Statistics},
  publisher = {Institute of Mathematical Statistics},
  pages={461--464},
  volume = {6},
  year={1978},
  URL = {http://www.jstor.org/stable/2958889}
}

@article{shi2021model,
  title={{Model-Robust Subdata Selection for Big Data}},
  author={Shi, Chenlu and Tang, Boxin},
  journal={Journal of Statistical Theory and Practice},
  volume={15},
  number={4},
  pages={1--17},
  year={2021},
  publisher={Springer},
  doi={https://doi.org/10.1007/s42519-021-00217-9}
}

@book{van2000asymptotic,
  title={{Asymptotic Statistics}},
  author={Van der Vaart, Aad W},
  volume={3},
  year={2000},
  isbn={9780521784504},
  lccn={98015176},
  series={Asymptotic Statistics},
  publisher={Cambridge university press}
}

@article{vaughan2020efficient,
  title={{Efficient big data model selection with applications to fraud detection}},
  author={Vaughan, Gregory},
  journal={International Journal of Forecasting},
  volume={36},
  number={3},
  pages={1116--1127},
  year={2020},
  issn = {0169-2070},
  publisher={Elsevier},
  doi={https://doi.org/10.1016/j.ijforecast.2018.03.002}
}

@article{wang2016statistical,
  title={{Statistical methods and computing for big data}},
  author={Wang, Chun and Chen, Ming-Hui and Schifano, Elizabeth and Wu, Jing and Yan, Jun},
  journal={Statistics and Its Interface},
  volume={9},
  number={4},
  pages={399-414},
  year={2016},
  publisher={International Press of Boston},
  doi={https://doi.org/10.4310/SII.2016.v9.n4.a1}
}

@article{wang2019more,
  title={{More Efficient Estimation for Logistic Regression with Optimal Subsamples}},
  author={Wang, HaiYing},
  journal={Journal of Machine Learning Research},
  volume={20},
  number={132},
  pages={1--59},
  url = {http://jmlr.org/papers/v20/18-596.html},
  year={2019}
}

@article{wang2019linear,
  title={{Information-Based Optimal Subdata Selection for Big Data Linear Regression}},
  author={Wang, HaiYing and Yang, Min and Stufken, John},
  journal={Journal of the American Statistical Association},
  volume={114},
  number={525},
  pages={393--405},
  year={2019},
  publisher={Taylor \& Francis},
  doi={https://doi.org/10.1080/01621459.2017.1408468}
}

@article{wang2018logistic,
  title={{Optimal Subsampling for Large Sample Logistic Regression}},
  author={Wang, HaiYing and Zhu, Rong and Ma, Ping},
  journal={Journal of the American Statistical Association},
  volume={113},
  number={522},
  pages={829--844},
  year={2018},
  publisher={Taylor \& Francis},
  doi={https://doi.org/10.1080/01621459.2017.1292914}
}

@article{xue2020online,
  title={{An online updating approach for testing the proportional hazards assumption with streams of survival data}},
  author={Xue, Yishu and Wang, HaiYing and Yan, Jun and Schifano, Elizabeth D},
  journal={Biometrics},
  volume={76},
  number={1},
  pages={171--182},
  year={2020},
  publisher={Wiley Online Library},
  doi={https://doi.org/10.1111/biom.13137}
}

@article{yang2014modeling,
  title={{Modeling Taxi Trip Demand by Time of Day in New York City}},
  author={Yang, Ci and Gonzales, Eric J},
  journal={Transportation Research Record},
  volume={2429},
  number={1},
  pages={110--120},
  year={2014},
  publisher={SAGE Publications Sage CA: Los Angeles, CA},
  doi={https://doi.org/10.3141/2429-12}
}

@inbook{yang2017modeling,
  title={{Modeling Taxi Demand and Supply in New York City Using Large-Scale Taxi GPS Data}},
  author={Yang, Ci and Gonzales, Eric J},
  booktitle={Seeing Cities Through Big Data: Research, Methods and Applications in Urban Informatics},
  pages={405--425},
  year={2017},
  isbn={978-3-319-40902-3},
  publisher={Springer International Publishing},
  doi={https://doi.org/10.1007/978-3-319-40902-3_22}
}

@article{yao2019softmax,
  title={{Optimal subsampling for softmax regression}},
  author={Yao, Yaqiong and Wang, HaiYing},
  journal={Statistical Papers},
  volume={60},
  number={2},
  pages={585--599},
  year={2019},
  publisher={Springer},
  doi={https://doi.org/10.1007/s00362-018-01068-6}
}

@article{yao2021review,
  title={{A Review on Optimal Subsampling Methods for Massive Datasets}},
  author={Yao, Yaqiong and Wang, HaiYing},
  journal={Journal of Data Science},
  volume={19},
  number={1},
  pages={151--172},
  year={2021},
  publisher={School of Statistics, Renmin University of China},
  doi={https://doi.org/10.6339/21-JDS999}
}

@inproceedings{yazici2013big,
  title={{A big data driven model for taxi drivers' airport pick-up decisions in new york city}},
  author={Yazici, M Anil and Kamga, Camille and Singhal, Abhishek},
  booktitle={2013 IEEE International Conference on Big Data},
  pages={37--44},
  year={2013},
  organization={IEEE},
  doi={https://doi.org/10.1109/BigData.2013.6691775}
}

@article{yu2022subdata,
  title={{Subdata selection algorithm for linear model discrimination}},
  author={Yu, Jun and Wang, HaiYing},
  journal={Statistical Papers},
  pages={1--24},
  year={2022},
  doi={https://doi.org/10.1007/s00362-022-01299-8},
  publisher={Springer}
}

@article{zhang2015astronomy,
  title={{Astronomy in the big data era}},
  author={Zhang, Yanxia and Zhao, Yongheng},
  journal={Data Science Journal},
  volume={14},
  year={2015},
  publisher={Ubiquity Press},
  doi={https://doi.org/10.5334/dsj-2015-011}
}

@article{Laura2022Optimal,
  author = {Laura Deldossi and Chiara Tommasi},
  title = {{Optimal design subsampling from Big Datasets}},
  journal = {Journal of Quality Technology},
  volume = {54},
  number = {1},
  pages = {93--101},
  year  = {2022},
  publisher = {Taylor \& Francis},
  doi = {https://doi.org/10.1080/00224065.2021.1889418}
}

%\renewcommand\bibname{\large \bf References}]
%% use bibfile 
%\bibliographystyle{chicago} % Chicago style, author-year citations
%\bibliography{bib}   % name your BibTeX data base

\begin{appendices}\label{Appendix}

\section{Proof of Theorem~\ref{The:5}}\label{Appendix:Theorem}

\begin{proof}
    \begin{align}
        & \sum_{q=1}^{Q} \alpha_q \mbox{tr}{(\bm{V}_q)} = \sum_{q=1}^{Q} \alpha_q \mbox{tr}\Big( \bm{J}^{-1}_{\bm{X}_q} {\bm{V}_q}_c \bm{J}^{-1}_{\bm{X}_q} \Big) \notag \\
        = & \frac{1}{N^2 r} \sum_{q=1}^{Q} \alpha_q \sum_{i=1}^{N} \mbox{tr}\Big[\frac{1}{{\phi_q}_i} \{y_i - \dot{\psi}(u(\hat{\bm{\theta}_q}^T_{MLE}\bm{x}_{qi}))\}^2 \bm{J}^{-1}_{\bm{X}_q} \notag \\ & \hspace{3cm} \dot{u}( \hat{\bm{\theta}_q}^T_{MLE} \bm{x}_{qi}) \bm{x}_{qi} [\dot{u}(\hat{\bm{\theta}_q}_{MLE}^T \bm{x}_{qi})\bm{x}_{qi}]^T \bm{J}^{-1}_{\bm{X}_q} \Big] \notag \\
        = & \frac{1}{N^2 r} \sum_{q=1}^{Q} \alpha_q \sum_{i=1}^{N} \Big[\frac{1}{{\phi_q}_i} \{y_i - \dot{\psi}(u(\hat{\bm{\theta}_q}^T_{MLE} \bm{x}_{qi}))\}^2 || \bm{J}^{-1}_{\bm{X}_q} \dot{u}(\hat{\bm{\theta}_q}^T_{MLE} \bm{x}_{qi}) \bm{x}_{qi} ||^2 \Big] \notag \\
        = & \frac{1}{N^2 r} \sum_{q=1}^{Q} \alpha_q \Big(\sum_{i=1}^{N} {\phi_q}_i \Big) \sum_{i=1}^{N} \Big[ \phi^{-1}_{qi} \{y_i - \dot{\psi} (u(\hat{\bm{\theta}_q}^T_{MLE} \bm{x}_{qi}))\}^2 || \bm{J}^{-1}_{\bm{X}_q} \dot{u}( \hat{\bm{\theta}_q}^T_{MLE} \bm{x}_{qi}) \bm{x}_{qi} ||^2 \Big] \notag \\
        = & \frac{\alpha_1}{N^2 r}  \Big(\sum_{i=1}^{N} {\phi_1}_i \Big) \sum_{i=1}^{N} \Big[ \phi^{-1}_{1i} \{y_i - \dot{\psi} (u(\hat{\bm{\theta}_1}^T_{MLE} \bm{x}_{1i}))\}^2 || \bm{J}^{-1}_{\bm{X}_1} \dot{u}( \hat{\bm{\theta}_1}^T_{MLE} \bm{x}_{1i}) \bm{x}_{1i} ||^2 \Big] + \ldots + \notag \\ & \frac{\alpha_Q}{N^2 r} \Big(\sum_{i=1}^{N} {\phi_Q}_i \Big) \sum_{i=1}^{N} \Big[ \phi^{-1}_{Qi} \{y_i - \dot{\psi} (u( \hat{\bm{\theta}_Q}^T_{MLE} \bm{x}_{Qi}))\}^2 || \bm{J}^{-1}_{\bm{X}_Q} \dot{u}(\hat{\bm{\theta}_Q}^T_{MLE} \bm{x}_{Qi}) \bm{x}_{Qi}||^2 \Big]\notag \\
        \ge & \frac{\alpha_1}{N^2 r} \sum_{i=1}^{N} \{y_i - \dot{\psi} (u(\hat{\bm{\theta}_1}^T_{MLE} \bm{x}_{1i}))\}^2 || \bm{J}^{-1}_{\bm{X}_1} \dot{u}(\hat{\bm{\theta}_1}^T_{MLE} \bm{x}_{1i}) \bm{x}_{1i} ||^2 \Big] + \ldots + \notag \\ & \frac{\alpha_Q}{N^2 r} \sum_{i=1}^{N} \{y_i - \dot{\psi} (u(\hat{\bm{\theta}_Q}^T_{MLE} \bm{x}_{Qi}))\}^2 || \bm{J}^{-1}_{\bm{X}_Q} \dot{u}(\hat{\bm{\theta}_Q}^T_{MLE} \bm{x}_{Qi}) \bm{x}_{Qi} ||^2 \Big] \notag \\
        = & \frac{1}{N^2 r} \sum_{q=1}^{Q} \alpha_q \sum_{i=1}^{N} \{y_i - \dot{\psi} (u(\hat{\bm{\theta}_q}^T_{MLE}  \bm{x}_{qi}))\}^2 || \bm{J}^{-1}_{\bm{X}_q} \dot{u}( \hat{\bm{\theta}_q}^T_{MLE} \bm{x}_{qi}) \bm{x}_{qi} ||^2 \Big]
    \end{align}
    
    where the last inequality follows from the Cauchy-Schwarz inequality, and the equality in it holds if and only if 
    %$${\phi_q}_i \propto |y_i - \dot{\psi} (u(\hat{\bm{\theta}_q}^T_{MLE} \bm{x}_{qi}))| \, ||\bm{J}^{-1}_{\bm{X_q}} \bm{x}_{qi}|| \, I\{|y_i - \dot{\psi} (u(\hat{\bm{\theta}_q}^T_{MLE} \bm{x}_{qi}))| \}, ||\bm{J}^{-1}_{\bm{X_q}} \bm{x}_{qi}|| > 0 \}. $$
    
    $${\phi_q}_i \propto |y_i - \dot{\psi} (u(\hat{\bm{\theta}_q}^T_{MLE} \bm{x}_{qi}))| \, || \bm{J}^{-1}_{\bm{X}_q} \dot{u}( \hat{\bm{\theta}_q}^T_{MLE} \bm{x}_{qi}) \bm{x}_{qi} ||.$$
    
    Here we define $0/0=0$, and this equivalent to removing data points with $|y_i - \dot{\psi} (u(\hat{\bm{\theta}_q}^T_{MLE} \bm{x}_{qi}))| = 0$ in the expression of ${\bm{V}_q}_c$.
    
    For the $q$-th model sub-sampling probabilities would be 
    \begin{equation}
        {\phi_q}^{mMSE}_i = \frac{|y_i - \dot{\psi}(u( \hat{\bm{\theta}_q}^T_{MLE} \bm{x}_{qi}))|\, || \bm{J}^{-1}_{\bm{X}_q} \dot{u}(\hat{\bm{\theta}_q}^T_{MLE} \bm{x}_{qi}) \bm{x}_{qi} ||}{\sum_{j=1}^{N} |y_j - \dot{\psi}(u(\hat{\bm{\theta}_q}^T_{MLE} \bm{x}_{qj}))|\, || \bm{J}^{-1}_{\bm{X}_q} \dot{u}( \hat{\bm{\theta}_q}^T_{MLE} \bm{x}_{qj}) \bm{x}_{qj} ||},
    \end{equation}
    $i=1,\ldots,N$ and  $\phi_q = \sum_{i=1}^{N} {\phi_q}_i^{mMSE} = 1$.
    
    If so, using the {\it a priori} probabilities $\alpha_q$ the model average optimal sub-sampling probabilities is chosen such that 
    \begin{equation}
        {\phi_i}^{mMSE} = \sum_{q=1}^{Q} \alpha_q {\phi_q}^{mMSE}_i,
    \end{equation}
    $i=1,\ldots,N$, $\sum_{q=1}^{Q} \alpha_q = 1$ then $\sum_{q=1}^{Q} \alpha_q \mbox{tr}(\bm{V}_q)$ attains its minimum. \\
    
    Similarly optimal sub-sampling probabilities can be obtained for $L$-optimality.
\end{proof}

\section{Algorithms}\label{Appendix:Algorithms}

\begin{algorithm}[H]
\SetAlgoLined

    \nonl \textbf{Stage 1} \\ \vspace{0.1cm}
    \textbf{Random Sampling:} Assign $\bm{\phi} = (\phi_1,\ldots,\phi_N)$ for $F_N$. Here, $\bm{\phi}=\phi^{prop}$, where $\phi^{prop}_i=(2N_0)^{-1}$ if the $i$-th response element is zero, otherwise $\phi^{prop}_i=(2N_1)^{-1}$, and $N_0$ and $N_1$ are the number of elements in $\bm{y}$ when $y_i \in 0$ and $y_i \in 1$, respectively.\\ \nonl 
    According to $\bm{\phi}$ draw a random sub-sample of size $r_0$, $S_{r_0}=\{\bm{x}^{r_0}_l,y^{r_0}_l,\phi^{r_0}_l\}_{l=1}^{r_0}$. \\
    \textbf{Estimation:} Based on $S_{r_0}$, using the Newton Raphson method find $\tilde{\bm{\theta}}^{r_0}$ until $\tilde{\bm{\theta}}^{t+1}$ and $\tilde{\bm{\theta}}^t$ are `close' enough ($\tilde{\bm{\theta}}^{t+1} - \tilde{\bm{\theta}}^t < 10^{-4}$),
    \begin{equation}
    \begin{aligned}
        \tilde{\bm{\theta}}^{t+1} = \tilde{\bm{\theta}}^t + \Bigg(\sum_{l=1}^{r_0} \frac{\pi_l(\tilde{\bm{\theta}}^t)(1-\pi_l(\tilde{\bm{\theta}}^t))\bm{x}^{r_0}_l (\bm{x}^{r_0}_l)^T}{\phi^{r_0}_l}\Bigg)^{-1} \Bigg(\sum_{l=1}^{r_0} \frac{(y^{r_0}_l-\pi_l(\tilde{\bm{\theta}}^t))\bm{x}^{r_0}_l}{\phi^{r_0}_l} \Bigg), \notag
    \end{aligned}    
    \end{equation}
    where $\mbox{logit} ~ \pi_l(\bm{\theta})=\bm{\theta}^T\bm{x}_l$. \nonl \\  
    \vspace{0.1cm} 
    \textbf{Stage 2} \\
    \vspace{0.1cm}
    \textbf{Optimal sub-sampling probability:} Using $\tilde{\bm{\theta}}^{r_0}$ estimate optimal sub-sampling probabilities, 
    \begin{equation}
    \begin{aligned}
        \phi^{mMSE}_i= \frac{|y_i-\pi_i(\tilde{\bm{\theta}}^{r_0})|\,||
        \tilde{\bm{J}}^{-1}_{\bm{X}} \bm{x}_i||}{ \sum_{j=1}^N |y_j-\pi_j(\tilde{\bm{\theta}}^{r_0})|\,|| \tilde{\bm{J}}^{-1}_{\bm{X}} \bm{x}_j||} \quad \mbox{or} \quad \phi^{mV_c}_i=  \frac{|y_i-\pi_i(\tilde{\bm{\theta}}^{r_0})|\,||\bm{x}_i||}{ \sum_{j=1}^N |y_j-\pi_j(\tilde{\bm{\theta}}^{r_0})|\,||\bm{x}_j||}, \notag
    \end{aligned}
    \end{equation}
    where $\tilde{\bm{J_X}}={(Nr_0)}^{-1} \sum_{l=1}^{r_0} (\phi_l^{r_0})^{-1} \pi_l(\tilde{\bm{\theta}}^{r_0})(1-\pi_l(\tilde{\bm{\theta}}^{r_0})) {\bm{x}_l}^{r_0} ({\bm{x}_l}^{r_0})^T$ and $i=1,\ldots,N$.\\
    \textbf{Optimal sub-sampling and estimation:} Using estimated $\bm{\phi}^{mMSE}$ or $\bm{\phi}^{mV_c}$ draw a sub-sample $S_r$ completely at random (with replacement) of size $r$ from $F_N$, such that $S_r=\{\bm{x}^r_l,y^r_l,\phi^r_l\}_{l=1}^r$. \\ \nonl 
    Form $S_{r_0+r}$ by combining $S_{r_0}$, $S_r$ and obtain $\tilde{\bm{\theta}}$ until $\tilde{\bm{\theta}}^{t+1}$ and $\tilde{\bm{\theta}}^t$ are 'close' enough ($\tilde{\bm{\theta}}^{t+1} - \tilde{\bm{\theta}}^t < 10^{-4}$):
    \begin{equation}
    \begin{aligned}
        \tilde{\bm{\theta}}^{t+1} = \tilde{\bm{\theta}}^t + \Bigg(\sum_{k \in \{r_0,r\}} \sum_{l=1}^{k} \frac{\pi_l(\tilde{\bm{\theta}}^t)(1-\pi_l(\tilde{\bm{\theta}}^t))\bm{x}^k_l (\bm{x}^k_l)^T}{\phi^k_l}\Bigg)^{-1} \Bigg(\sum_{k \in \{r_0,r\}} \sum_{l=1}^{k} \frac{(y^k_l-\pi_l(\tilde{\bm{\theta}}^t))\bm{x}^k_l}{\phi^k_l} \Bigg). \notag
    \end{aligned}
    \end{equation} \\ \nonl 
    Estimate the variance-covariance matrix of $\tilde{\bm{\theta}}$, $\tilde{\bm{V}}$, by $\tilde{\bm{J}}^{-1}_{\bm{X}} \tilde{\bm{V}}_c\tilde{\bm{J}}^{-1}_{\bm{X}}$, where,
    \begin{align}
        \tilde{\bm{J_X}} = & \frac{1}{N(r_0+r)} \sum_{k \in \{r_0,r\}} \sum_{l=1}^{k} \frac{\pi_l(\tilde{\bm{\theta}})(1-\pi_l(\tilde{\bm{\theta}})) {\bm{x}^k_l} ({\bm{x}^k_l})^T}{\phi^k_l} \notag \\ \mbox{and} &\quad \tilde{\bm{V_c}} = \frac{1}{N^2(r_0+r)^2} \sum_{k \in \{r_0,r\}} \sum_{l=1}^{k} \frac{(y^k_l-{\pi_l}(\tilde{\bm{\theta}}))^2 {\bm{x}^k_l} ({\bm{x}^k_l})^T}{ ({\phi^k_l})^2}. \notag
    \end{align} \\
    \textbf{Output :} $\tilde{\bm{\theta}}, S_{r_0+r}$ and $\tilde{\bm{V}}$.
\caption{Two stage optimal sub-sampling algorithm for logistic regression.} \label{Algo:OSMAC}
\end{algorithm}
\newpage
\begin{algorithm}[H]
\SetAlgoLined

    \nonl \textbf{Stage 1} \\ \vspace{0.1cm}
    \textbf{Random Sampling:} Assign $\bm{\phi} = (\phi_1,\ldots,\phi_N)$ for $F_N$. Here, $\phi_i = {N^{-1}}$.\\ \nonl 
    According to $\bm{\phi}$ draw a random sub-sample of size $r_0$, $S_{r_0}=\{\bm{x}^{r_0}_l,y^{r_0}_l,\phi^{r_0}_l\}_{l=1}^{r_0} = (\bm{X}^{r_0},\bm{y}^{r_0},\bm{\phi}^{r_0})$.\\
    \textbf{Estimation:} Based on $S_{r_0}$, find:
    \begin{equation}
    \begin{aligned}
        \tilde{\bm{\theta}}^{r_0} = \argmaxA_{\bm{\theta}} ~ \log{ L(\bm{\theta}|\bm{X}^{r_0},\bm{y}^{r_0},\bm{\phi}^{r_0})} \equiv  \argmaxA_{\bm{\theta}} ~ \frac{1}{r_0} \sum_{l=1}^{r_0} \Big[\frac{y^{r_0}_l \bm{\theta}^T \bm{x}^{r_0}_l - \lambda_l(\bm{\theta}) - \log{(y^{r_0}_l!)}}{\phi^{r_0}_l} \Big], \notag
    \end{aligned}
    \end{equation}
     where $\log~{\lambda_l(\bm{\theta})} = \bm{\theta}^T\bm{x}_l.$ \nonl \\  
    \vspace{0.1cm} 
    \textbf{Stage 2} \\
    \vspace{0.1cm}
    \textbf{Optimal sub-sampling probability:} Using $\tilde{\bm{\theta}}^{r_0}$ estimate optimal sub-sampling probabilities, 
    \begin{equation}
    \begin{aligned}
        \phi^{mMSE}_i= \frac{|y_i-\lambda_i(\tilde{\bm{\theta}}^{r_0})| \, || \tilde{\bm{J}}^{-1}_{\bm{X}} \bm{x}_i||}{ \sum_{j=1}^N |y_j-\lambda_j(\tilde{\bm{\theta}}^{r_0})| \, || \tilde{\bm{J}}^{-1}_{\bm{X}} \bm{x}_j||} \quad \mbox{or} \quad \phi^{mV_c}_i=  \frac{|y_i-\lambda_i(\tilde{\bm{\theta}}^{r_0})| \, ||\bm{x}_i||}{ \sum_{j=1}^N |y_j-\lambda_j(\tilde{\bm{\theta}}^{r_0})| \, || \bm{x}_j||}, \notag
    \end{aligned}
    \end{equation}
    where $\tilde{\bm{J_X}}={(Nr_0)}^{-1} \sum_{l=1}^{r_0} (\phi_l^{r_0})^{-1} \lambda_l(\tilde{\bm{\theta}}^{r_0}) {\bm{x}_l}^{r_0} ({\bm{x}_l}^{r_0})^T$ and $i=1,\ldots,N$.\\
    \textbf{Optimal sub-sampling and estimation:} Using estimated $\bm{\phi}^{mMSE}$ or $\bm{\phi}^{mV_c}$ draw a sub-sample $S_r$ completely at random (with replacement) of size $r$ from $F_N$, such that $S_r=\{\bm{x}^r_l,y^r_l,\phi^r_l\}_{l=1}^r = (\bm{X}^r,\bm{y}^r,\bm{\phi}^r)$. \\ \nonl 
    Form $S_{r_0+r}$ by combining $S_{r_0}$, $S_r$ and obtain: 
    \begin{align}
        \tilde{\bm{\theta}} = & \argmaxA_{\bm{\theta}} ~ \log{ L(\bm{\theta}|\bm{X}^{r_0},\bm{y}^{r_0},\bm{\phi}^{r_0},\bm{X}^r,\bm{y}^r,\bm{\phi}^r)} \notag \\ \equiv & \argmaxA_{\bm{\theta}} ~\frac{1}{r_0+r} \sum_{k \in \{r_0,r\}} \sum_{l=1}^{k} \Big[\frac{y^k_l \bm{\theta}^T\bm{x}^k_l  - \lambda_l(\bm{\theta}) - \log{(y^k_l!)}}{\phi^k_l} \Big]. \notag
    \end{align} \\ \nonl 
    Estimate the variance-covariance matrix of $\tilde{\bm{\theta}}$, $\tilde{\bm{V}}$, by $\tilde{\bm{J}}^{-1}_{\bm{X}} \tilde{\bm{V}}_c \tilde{\bm{J}}^{-1}_{\bm{X}}$, where,
    \begin{align}
        \tilde{\bm{J_X}} = & \frac{1}{N(r_0+r)} \sum_{k \in \{r_0,r\}} \sum_{l=1}^{k} \frac{\lambda_l(\tilde{\bm{\theta}}) {\bm{x}^k_l} ({\bm{x}^k_l})^T}{ \phi^k_l} \notag \\ \mbox{and} & \quad \tilde{\bm{V_c}} = \frac{1}{N^2(r_0+r)^2} \sum_{k \in \{r_0,r\}} \sum_{l=1}^{k} \frac{(y^k_l-{\lambda_l}(\tilde{\bm{\theta}}))^2 {\bm{x}^k_l} ({\bm{x}^k_l})^T}{ ({\phi^k_l})^2}. \notag
    \end{align} \\
    \textbf{Output :} $\tilde{\bm{\theta}}, S_{r_0+r}$ and $\tilde{\bm{V}}$.
\caption{Two stage optimal sub-sampling algorithm for Poisson regression.}\label{Algo:OSPAC}
\end{algorithm}
\newpage
\begin{algorithm}[H]
\SetAlgoLined

    \nonl \textbf{Stage 1} \\ \vspace{0.1cm}
    \textbf{Random Sampling:} Assign $\bm{\phi} = (\phi_1,\ldots,\phi_N)$ for $F_{N}$. Here, $\bm{\phi} =\phi^{prop}$, where $\phi^{prop}_i=(2N_0)^{-1}$ if the $i$-th response element is $0$, otherwise $\phi^{prop}_i=(2N_1)^{-1}$, and $N_0$ and $N_1$ are the number of elements in $\bm{y}$ when $y_i \in 0$ and $y_i \in 1$, respectively.\\ \nonl 
    According to $\bm{\phi}$ draw random sub-samples of size $r_0$, such that  $S_{r_0}=\{h_q(\bm{{x_0}}^{r_0}_{l}),y^{r_0}_l,\phi^{r_0}_l\}_{l=1}^{r_0}$ for $q=1,\ldots,Q$. \\
    \textbf{Estimation:} For $q=1,\ldots,Q$ and $S_{r_0}$, using the Newton Raphson method find $\tilde{\bm{\theta}_q}^{r_0}$ until $\tilde{\bm{\theta}_q}^{t+1}$ and $\tilde{\bm{\theta}_q}^t$ are `close' enough ($\tilde{\bm{\theta}_q}^{t+1} - \tilde{\bm{\theta}_q}^t < 10^{-4}$),
    \begin{equation}
    \begin{aligned}
        \tilde{\bm{\theta}_q}^{t+1} = \tilde{\bm{\theta}_q}^t + \Bigg(\sum_{l=1}^{r_0} \frac{\pi_l(\tilde{\bm{\theta}_q}^t)(1-\pi_l(\tilde{\bm{\theta}_q}^t)) \bm{x}_{ql} (\bm{x}_{ql})^T }{ \phi_l} \Bigg)^{-1} \Bigg(\sum_{l=1}^{r_0} \frac{(y_l - \pi_l(\tilde{\bm{\theta}_q}^t)) \bm{x}_{ql}}{\phi_l} \Bigg), \notag
    \end{aligned}
    \end{equation}
     where $\mbox{logit} ~ \pi_l(\bm{\theta}_q)= \bm{\theta}^T_q\bm{x}_{ql}$. \nonl \\  
    \vspace{0.1cm} 
    \textbf{Stage 2} \\
    \vspace{0.1cm}
    \textbf{Optimal sub-sampling probability:} Based on $\tilde{\bm{\theta}_q}^{r_0}$ and $\alpha_q$ (for $q=1,\ldots,Q$) estimate optimal sub-sampling probabilities, $$\phi^{mMSE}_i= \sum_{q=1}^{Q} \frac{\alpha_q |y_i-\pi_i(\tilde{\bm{\theta}_q}^{r_0})| \, || \tilde{\bm{J}}^{-1}_{\bm{X}_q} \bm{x}_{qi} ||}{ \sum_{j=1}^N |y_j-\pi_j(\tilde{\bm{\theta}_q}^{r_0})| \, || \tilde{\bm{J}}^{-1}_{\bm{X}_q} \bm{x}_{qj} ||} \quad \mbox{or} \quad \phi^{mV_c}_i= \sum_{q=1}^{Q} \frac{\alpha_q |y_i-\pi_i(\tilde{\bm{\theta}_q}^{r_0})| \, || \bm{x}_{qi} ||}{ \sum_{j=1}^N |y_j-\pi_j(\tilde{\bm{\theta}_q}^{r_0})| \, || \bm{x}_{qj}||},$$ where $\tilde{\bm{J}}_{\bm{X}_q}={(Nr_0)}^{-1} \sum_{l=1}^{r_0} (\phi_l^{r_0})^{-1} \pi_l (\tilde{\bm{\theta}_q}^{r_0})(1-\pi_l(\tilde{\bm{\theta}_q}^{r_0})) {\bm{x}^{r_0}_{ql}} (\bm{x}^{r_0}_{ql})^T$ and $i=1,\ldots,N$.\\
    \textbf{Optimal sub-sampling and estimation:} Based on $\bm{\phi}^{mMSE}$ or $\bm{\phi}^{mV_c}$ draw a sub-sample of size $r$ from $F_{N}$, such that $S_r=\{h_q(\bm{{x_0}}^r_{l}),y^r_l,\phi^r_l\}_{l=1}^r$ for $q=1,\ldots,Q$. \\ \nonl 
    Combine $S_{r_0},S_r$ and form $S_{(r_0+r)}$, and obtain $\tilde{\bm{\theta}_q}$ until $\tilde{\bm{\theta}_q}^{t+1}$ and $\tilde{\bm{\theta}_q}^t$ are 'close' enough ($\tilde{\bm{\theta}_q}^{t+1} - \tilde{\bm{\theta}_q}^t < 10^{-4}$):
    \begin{equation}
    \begin{aligned}
        & \tilde{\bm{\theta}_q}^{t+1} = \tilde{\bm{\theta}_q}^t + \Bigg[\sum_{k \in \{r_0,r\}} \sum_{l=1}^{k} \frac{\pi_l(\tilde{\bm{\theta}_q}^t)(1-\pi_l(\tilde{\bm{\theta}_q}^t)) \bm{x}^{k}_{ql}  (\bm{x}^{k}_{ql})^T}{\phi^k_l}\Bigg]^{-1} \Bigg[\sum_{k \in \{r_0,r\}}\sum_{l=1}^{k} \frac{(y^k_l-\pi_l(\tilde{\bm{\theta}_q}^t)) \bm{x}^{k}_{ql} }{\phi^k_l} \Bigg]. \notag
    \end{aligned}
    \end{equation} \\ \nonl 
    Estimate the variance-covariance matrix of $\tilde{\bm{\theta}}_q$, $\tilde{\bm{V}}_q$, by $\tilde{\bm{J}}^{-1}_{\bm{X}_q} \tilde{\bm{V}}_{q_c} \tilde{\bm{J}}^{-1}_{\bm{X}_q}$, where,
    \begin{align}
        \tilde{\bm{J}}_{\bm{X}_q} = & \frac{1}{N(r_0+r)} \sum_{k\in \{r_0,r\}} \sum_{l=1}^{k} \frac{\pi_l(\tilde{\bm{\theta}_q})(1-\pi_l(\tilde{\bm{\theta}_q})) \bm{x}^{k}_{ql} (\bm{x}^{k}_{ql})^T}{ \phi^k_l} \notag \\ \mbox{and} & \quad \tilde{\bm{V}}_{q_c} = \frac{1}{N^2(r_0+r)^2} \sum_{k \in \{r_0,r\}} \sum_{l=1}^{k} \frac{(y^k_l-{\pi_l}(\tilde{\bm{\theta}_q}))^2 \bm{x}^{k}_{ql} (\bm{x}^{k}_{ql})^T}{ ({\phi^k_l})^2}. \notag
    \end{align} \\ 
    \textbf{Output :} $\tilde{\bm{\theta}}_q, S_{(r_0+r)}$ and $\tilde{\bm{V}}_q$ for $q=1,\ldots,Q$.
\caption{Two stage model robust optimal sub-sampling algorithm for logistic regression.}\label{Algo:OSMACMR}
\end{algorithm}
\newpage
\begin{algorithm}[H]
\SetAlgoLined

    \nonl \textbf{Stage 1} \\ \vspace{0.1cm}
    \textbf{Random Sampling :} Assign $\bm{\phi} = (\phi_1,\ldots,\phi_N)$ for $F_{N}$. Here, $\phi_i = {N^{-1}}$.\\ \nonl 
    According to $\bm{\phi}$ draw random sub-samples of size $r_0$, such that  $S_{r_0}=\{h_q({\bm{x}_0}^{r_0}_{l}),y^{r_0}_l,\phi^{r_0}_l\}_{l=1}^{r_0} = (h_q(\bm{X}_0^{r_0}),\bm{y}^{r_0},\bm{\phi}^{r_0})$ for $q=1,\ldots,Q$. \\ 
    \textbf{Estimation:} For $q=1,\ldots,Q$ and $S_{r_0}$ find:  
    \begin{equation}
    \begin{aligned}
        \tilde{\bm{\theta}_q}^{r_0} = \argmaxA_{\bm{\theta}_q} ~ \log{L(\bm{\theta}_q|  \bm{X}^{r_0}_{q},\bm{y}^{r_0},\bm{\phi}^{r_0})} \equiv \argmaxA_{\bm{\theta}_q} ~ \frac{1}{r_0} \sum_{l=1}^{r_0} \Big[\frac{y^{r_0}_l \bm{\theta}^T_q \bm{x}^{r_0}_{ql}  - \lambda_l(\bm{\theta}_q) - \log{(y^{r_0}_l!)}}{\phi^{r_0}_l} \Big], \notag
    \end{aligned}    
    \end{equation}
     where $\log{\lambda_l(\bm{\theta}_q)} = \bm{\theta}^T_q \bm{x}_{ql}.$ \nonl \\  
    \vspace{0.1cm} 
    \textbf{Stage 2} \\
    \vspace{0.1cm}
    \textbf{Optimal sub-sampling probability:} For $q=1,\ldots,Q$, using $\tilde{\bm{\theta}_{q}^{r_0}}$ estimate optimal sub-sampling probabilities, 
    \begin{equation}
    \begin{aligned}
        \phi^{mMSE}_i= \sum_{q=1}^{Q}\frac{\alpha_q|y_i-\lambda_i(\tilde{\bm{\theta}_{q}}^{r_0})| \, || \tilde{\bm{J}}^{-1}_{\bm{X}_q} \bm{x}_{qi} ||}{ \sum_{j=1}^N |y_j-\lambda_j(\tilde{\bm{\theta}_{q}}^{r_0})| \, || \tilde{\bm{J}}^{-1}_{\bm{X}_q} \bm{x}_{qj} ||} \quad \mbox{or} \quad \phi^{mV_c}_i= \sum_{q=1}^{Q} \frac{\alpha_q|y_i-\lambda_i(\tilde{\bm{\theta}_{q}}^{r_0})| \, || \bm{x}_{qi} ||}{ \sum_{j=1}^N |y_j-\lambda_j(\tilde{\bm{\theta}_{q}}^{r_0})| \, || \bm{x}_{qj} ||}, \notag
    \end{aligned}    
    \end{equation}
    where $\tilde{\bm{J}}_{\bm{X}_q}={(Nr_0)}^{-1} \sum_{l=1}^{r_0} (\phi_l^{r_0})^{-1} \lambda_l(\tilde{\bm{\theta}_{q}}^{r_0}) \bm{x}^{r_0}_{ql} (\bm{x}^{r_0}_{ql})^T$ and $i=1,\ldots,N$.\\
    \textbf{Optimal sub-sampling and estimation:} Based on $\bm{\phi}^{mMSE}$ or $\bm{\phi}^{mV_c}$ draw a sub-sample of size $r$ from $F_{N}$, such that $S_r=\{h_q({\bm{x}_0}^r_{l}),y^r_l,\phi^r_l\}_{l=1}^r = (h_q(\bm{X}_0^r),\bm{y}^r_q, \bm{\phi}^r_q)$ for $q=1,\ldots,Q$.\\ \nonl 
    Combine $S_{r_0}$, $S_r$ and form $S_{(r_0+r)}$, and obtain: 
    \begin{align}
      \tilde{\bm{\theta}}_q = & \argmaxA_{\bm{\theta}_q} ~ \log{L(\bm{\theta}_q| \bm{X}^{r_0}_q, \bm{y}^{r_0}_q, \bm{\phi}^{r_0}_q,  \bm{X}^r_q,\bm{y}^r_q, \bm{\phi}^r_q)} \notag \\
      \equiv & \argmaxA_{\bm{\theta}_q} ~ \frac{1}{r_0+r} \sum_{k \in \{r_0,r\}} \sum_{l=1}^{k} \Big[\frac{y^k_l \bm{\theta}^T_q \bm{x}^k_{ql}  - \lambda_l(\bm{\theta}_q) - \log{(y^k_l!)}}{\phi^k_l} \Big]. \notag
    \end{align} \\ \nonl 
     Estimate the variance-covariance matrix of $\tilde{\bm{\theta}}_q$, $\tilde{\bm{V}}_q$, by $\tilde{\bm{J}}^{-1}_{\bm{X}_q} \tilde{\bm{V}}_{q_c} \tilde{\bm{J}}^{-1}_{\bm{X}_q}$, here,
    \begin{align}
        \tilde{\bm{J}}_{\bm{X}_q} = & \frac{1}{N(r_0+r)}  \sum_{k \in \{r_0,r\}} \sum_{l=1}^{k} \frac{\lambda_l(\tilde{\bm{\theta}_q}) \bm{x}^k_{ql} (\bm{x}^k_{ql})^T}{ \phi^k_l}  \notag \\ \mbox{and} & \quad \tilde{\bm{V}}_{q_c} = \frac{1}{N^2(r_0+r)^2} \sum_{k \in \{r_0,r\}} \sum_{l=1}^{k} \frac{(y^k_l-{\lambda_l}(\tilde{\bm{\theta}_q}))^2 \bm{x}^k_{ql} (\bm{x}^k_{ql})^T}{ ({\phi^k_l})^2}. \notag
    \end{align} \\ 
    \textbf{Output :} $\tilde{\bm{\theta}}_q, S_{(r_0+r)}$ and $\tilde{\bm{V}}_q$ for $q=1,\ldots,Q$.
\caption{Two stage model robust optimal sub-sampling algorithm for Poisson regression.}\label{Algo:OSPACMR}
\end{algorithm}

\section{Code materials for the simulation setup and real world applications}\label{Appendix:Code}

\begin{enumerate}
    \item Logistic Regression
    \begin{enumerate}
        \item \href{https://github.com/Amalan-ConStat/RS_OS_MROS_LR_TwoVar_Exponential}{Two covariate model with covariate data generated by the Exponential distribution.}
        \item \href{https://github.com/Amalan-ConStat/RS_OS_MROS_LR_TwoVar_Normal}{Two covariate model with covariate data generated by the Normal distribution.}
        \item \href{https://github.com/Amalan-ConStat/RS_OS_MROS_LR_RWS_SkinData}{Skin segmentation data.}
    \end{enumerate}
    \item Poisson Regression
    \begin{enumerate}
        \item \href{https://github.com/Amalan-ConStat/RS_OS_MROS_PR_TwoVar_Normal}{Two covariate model with covariate data generated by the Normal distribution.}
        \item \href{https://github.com/Amalan-ConStat/RS_OS_MROS_PR_TwoVar_Uniform}{Two covariate model with covariate data generated by the Uniform distribution.}
        \item \href{https://github.com/Amalan-ConStat/RS_OS_MROS_PR_RWS_NYC_Taxi_Fare_2013_Full}{New York City taxi fare data.}
    \end{enumerate}
\end{enumerate}

\section{Figures}\label{Appendix:Figures}
\newpage
\begin{figure}[H]
    \centering
    \includegraphics[width=\textwidth,height=8.25in]{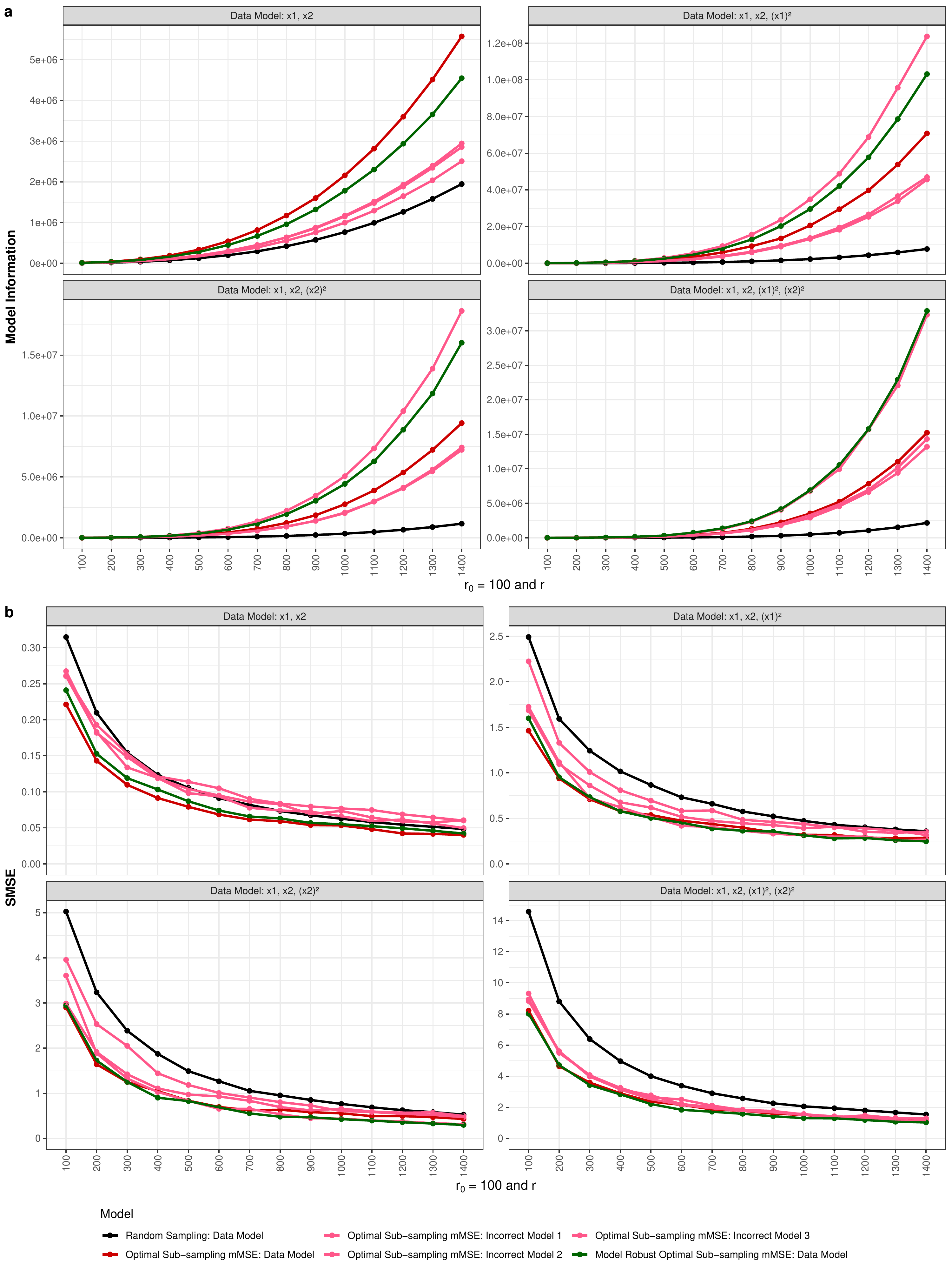} 
    \caption{(a) Model information (larger is better) and (b) SMSE (smaller is better) for the sub-sampling methods for the logistic regression model under $mMSE$. Covariate data were generated from the Exponential distribution.} \label{Fig:RWA_LR_TV_Exp}
\end{figure}

\begin{figure}[H]
    \centering
    \includegraphics[width=\textwidth,height=8.25in]{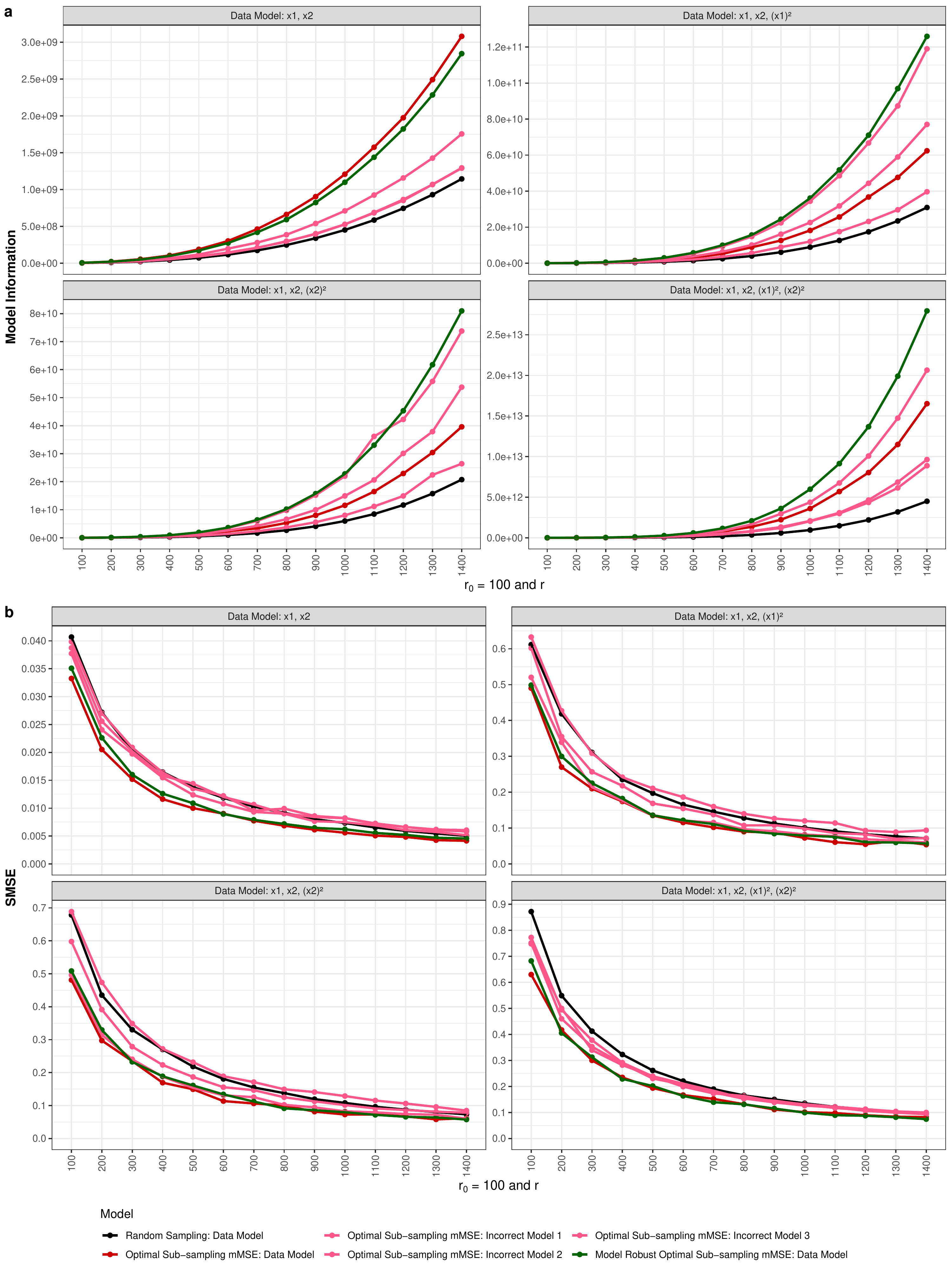}
    \caption{(a) Model information (larger is better) and (b) SMSE (smaller is better) for the sub-sampling methods for the Poisson regression model under $mMSE$. Covariate data were generated from the Uniform distribution.} \label{Fig:RWA_PR_TV_Uni}
\end{figure}

\end{appendices}

\end{document}